\documentclass[fleqn,useAMS,usenatbib]{mnras}
\usepackage{newtxtext,newtxmath}

\usepackage[T1]{fontenc}
\usepackage{ae,aecompl}
\usepackage{adjustbox}
\usepackage{graphicx}
\usepackage{epsfig}
\usepackage{amsmath}
\usepackage{amssymb}
\usepackage{natbib}
\usepackage{threeparttable} 
\usepackage{booktabs}
\usepackage{xcolor}
\usepackage{multirow}

\usepackage{subfigure}
\usepackage{multirow}

\usepackage{epstopdf}
\usepackage{times}

\newcommand{\swift}{\textit{Swift}}

\newcommand{\source}{Aql~X-1}

\def \nar {NewAR}

\def \mnras {MNRAS}
\def \apj {ApJ}
\def \apjs {ApJS}
\def \apjl {ApJL}
\def \aap {A\&A}
\def \nat {Nature}

\def \atel {ATel}

\title[Optical-UV/X-ray correlation in Aql X-1]{The connection between the UV/optical and X-ray emission in the neutron star low-mass X-ray binary Aql X-1}
\author[L\'{o}pez-Navas et al.]
{E. L\'{o}pez-Navas$^{1,2}$, N. Degenaar$^{1}$\thanks{e-mail: degenaar@uva.nl}, A.S. Parikh$^{1}$, J.V. Hern\'andez Santisteban$^{1,3}$, 
\newauthor J. van den Eijnden$^{1}$ \\
$^1$Anton Pannekoek Institute for Astronomy, University of Amsterdam, Science Park 904, 1098 XH, Amsterdam, the Netherlands\\
$^2$Instituto de F\'{i}sica y Astronom\'{i}a, Universidad de Valpara\'{i}so, Avda. Gran Breta\~{n}a 1111, Valpara\'{i}so, Chile\\
$^3$SUPA Physics and Astronomy, University of St Andrews, KY16 9SS Scotland, UK
}

\date{DRAFT VERSION}

\pubyear{2019}

\begin{document}
\label{firstpage}
\pagerange{\pageref{firstpage}--\pageref{lastpage}} 
\maketitle

\begin{abstract}
Accreting neutron stars and black holes in low-mass X-ray binaries (LMXBs) radiate across the electromagnetic spectrum. 
Linking the emission produced at different wavelengths can provide valuable information about the accretion process and any associated outflows. 
In this work, we study simultaneous X-ray and UV/optical observations of the neutron star LMXB \source, obtained with the \textit{Neil Gehrels Swift Observatory} during its 2013, 2014 and 2016 accretion outbursts. We find that the UV/optical and X-ray emission are strongly correlated during all three outbursts. For the 2013 and 2014 episodes, which had the best \textit{Swift} sampling, we find that the correlation between the UV/optical and X-ray fluxes is significantly steeper during the decay (soft state) of the outburst than during the rise (hard-to-soft state). We observe an UV/optical hysteresis behaviour that is likely linked to the commonly known X-ray spectral hysteresis pattern. For the decays of the three outbursts we obtain a correlation index that cannot be directly explained by any single model. We suspect that this is a result of multiple emission processes contributing to the UV/optical emission, but we discuss alternative explanations. Based on these correlations, we discuss which mechanisms are likely dominating the UV/optical emission of Aql X-1. 


\end{abstract}



\section{Introduction}
Our Galaxy contains $\sim$200 known low-mass X-ray binaries (LMXBs), in which a neutron star (NS) or a black hole (BH) is accreting gas from a late-type companion star that is less massive than the compact primary \citep{liu2006}. In nearly all LMXBs, the donor star overflows its Roche lobe and feeds gas into an accretion disc that spirals around the BH/NS. Thermal-viscous instabilities in this accretion disc cause many LMXBs to display transient behaviour; matter is rapidly accreted during outbursts when the gaseous disc is hot and ionised, while the accretion rate is strongly reduced during quiescent episodes, when matter in the disc is cold and recombined \citep[e.g.][for a review]{lasota01}. The brightness of LMXBs scales with the rate at which mass is accreted, causing these systems to be orders of magnitude brighter during outbursts than during quiescence.

LMXBs radiate across the electromagnetic spectrum. Their X-ray emission is typically ascribed to the (inner) accretion flow, while their emission from radio up to sub-mm arises from collimated outflows called jets \citep[e.g.][]{tetarenko2015,diaztrigo2018}. In the spectral region in between, i.e. in the ultra-violet (UV), optical and near-infrared (NIR) bands, multiple emission mechanisms may be operating. For instance, radiation at these wavelengths may be produced in the accretion disc, either due to viscous heating or due to irradiation \citep[][]{paradijs95,russell06}, in a hot flow \citep[at low accretion rates; e.g.][]{esin1997,shahbaz2003,veledina2013}, or in the jet \citep[e.g.][]{homan2005,russell06}. The donor star also radiates at optical and NIR wavelengths; this emission may be visible during quiescent episodes, but during outbursts the donor star is outshone by the other emission components \citep[e.g.][]{vanparadijs1994,charles2006}. Each of these different emission mechanisms are connected differently to the accretion flow. Studying correlations between the X-ray emission and the UV/optical/NIR emission, either on long or short time scales, can therefore give valuable insight into the  accretion process \citep[e.g.][]{russell06}.

Whereas the optical/NIR emission of LMXBs is routinely studied, UV studies are much sparser. This is primarily due to the fact that many LMXBs are located in the Galactic plane, and hence they suffer severely from interstellar extinction in the UV. For LMXBs with low Galactic extinction, however, UV studies can give additional information about the properties of the accretion flow \citep[e.g.][]{shrader1994,hynes1999,hynes2002,boroson2001,bernardini2013,boroson2014,froning2014}. One such LMXB with sufficiently low extinction to be detected in the UV band is \source.

\subsection{The Galactic neutron star LMXB \source}\label{subsec:source}
\source\ is a transient LMXB that was discovered over 5 decades ago \citep[][]{friedman1967}. The compact primary is known to be a NS because the source displays thermonuclear X-ray bursts \citep[e.g.][]{lewin1976,koyama1981}. The detection of burst oscillations \citep[][]{zhang1998} and a brief episode of coherent X-ray pulsations revealed that the NS is spinning at 550 Hz \citep[1.8~ms;][]{casella2008}. Rapid rotation rates on the order of hundreds of Hz are common for NS LMXBs \citep[see][for a recent overview of measured spin periods]{patruno2017_spin}, and are thought to be a result of the angular momentum transfer involved in the accretion process \citep[e.g.][]{alpar1982,wijnands1998}. The donor in \source\ is known to be a K star \citep[][]{thorstensen1978,matasanchez2017_aqlx1}, and the orbital period of the system is $\sim$ 19~hr \citep[][]{chevalier1991}.

Despite being a transient source, \source\ is particularly well characterised. This is in part due to the fact that it is one of the most frequently active transient LMXBs. It exhibits outbursts roughly once a year \citep[see e.g.][for a recent compilation of outbursts]{ootes2016}, and has been studied extensively for decades. Both optical/NIR and X-ray studies have shown that \source\ exhibits different types of outbursts: the classical fast rise and exponential decay (FRED) and the low-intensity state (LIS), where the optical-to-soft X-ray flux ratio is much higher than that seen during a FRED \citep[e.g.][]{maitra2008}. FRED outbursts have been categorized as long-high outbursts, medium-low and short-low outbursts depending on their duration and maximum flux \citep[e.g.][]{gungor2014_aqlx1,gungor2017_aqlx1}. It is not fully established what causes \source\ to have different outbursts, and if there are any physical differences between these classes. 

In this work, we investigate the connection between the UV/optical emission of \source, by studying simultaneous X-ray and UV/optical observations obtained with the \textit{Neil Gehrels Swift Observatory} \citep[\swift\ hereafter;][]{gehrels2004} during three well-monitored FRED outbursts. The aim of our study is to understand the origin of the UV/optical emission in this LMXB, and to investigate whether different types of FRED outbursts behave in the same way.


\section{Observations and data analysis}

\subsection{Selection of outbursts}\label{subsec:obs}
We searched the \swift\ data archive for outbursts of \source\ that had i) good sampling of the entire outburst with the X-ray Telescope (XRT) and ii) observations in one or more \textit{UV} filters on the UltraViolet and Optical Telescope (UVOT) consistently taken along the entire outburst. This yielded three different outbursts that were suitable for our analysis, those that occurred in 2013, 2014, and 2016. These outbursts have been categorised in the FRED class \citep{gungor2014_aqlx1, gungor2017_aqlx1}. The details of the observations analysed are listed in Table~\ref{table: outb}.

\begin{table}
\caption{Details of the outbursts analysed.}
\begin{center}
\resizebox{0.5\textwidth}{!}{%
\begin{threeparttable}
\begin{tabular}{cccccccc}
\hline\hline
Name\tnote{a}   & Year & Target ID & Observations &UVOT filters \\

\hline 

OUTB13 & 2013 & 00035323 & 004-026 &\textit{UM2,UW1,U,B,V}\\
& & 00032888 & 000-024\tnote{b} & \textit{UM2,UW1,U,B,V}\\
OUTB14 & 2014 & 00032888&026-045&\textit{UW2,UW1,U,B,V}\\
OUTB16 & 2016 & 00033665&074-089& \textit{UW2,U}\\
\hline
\end{tabular}

\begin{tablenotes}
  \item[a] Outburst indication adopted in this work.
  \item[b] Observations in the range 009-024 were discarded due to the low number of counts collected.
  \end{tablenotes}
  \end{threeparttable}
\label{table: outb}
}
\end{center}
\end{table}

\subsection{XRT data reduction and analysis}\label{subsec:xrt}
For each of our selected observations, we obtained the X-ray spectra and the associated response files using the online XRT data products tool\footnote{http://www.swift.ac.uk/user$\_$objects/}, which uses the latest version of the \textit{Swift} software and calibration \citep{evans09}. All WT spectra were grouped to require at least 20 counts per bin using the ftool \textsc{grppha} to ensure valid results using $\chi^{2}$ statistical analysis. 
We discarded all the PC mode spectra due to the low number of counts collected.

The XRT spectra were analysed using \textsc{Xspec} \citep[v. 12.10][]{xspec}. All spectral data were fitted with a simple model consisting of a power-law and a blackbody component (\textsc{powerlaw+bbodyrad}), affected by photoelectric absorption (\textsc{tbabs}). We assumed a constant column density ($N_{\mathrm{H}}$) of $0.36 \times 10^{22}~\mathrm{cm}^{-2}$, which was the best fit value obtained from fitting \textit{Suzaku} spectra (0.8--100 keV) when \source\ was in a soft state \citep{Sakurai2012}. This value is consistent with the hydrogen column density within our Galaxy in the direction of \source\ \citep[$N_{\mathrm{H}} = 0.31 \times 10^{22}~\mathrm{cm}^{-2}$;][]{HI4PI}. 

We note that when fitting all our XRT spectra simultaneously with the hydrogen column density free, we obtained typical fit values of $N_{\mathrm{H}} \simeq 0.5 \times 10^{22}~\mathrm{cm}^{-2}$. Whereas the inferred X-ray spectral parameters and fluxes do not differ much for this higher $N_{\mathrm{H}}$ value, it does lead to very different de-reddened UV fluxes (because the extinction correction affects the UV wavelengths more strongly). Since this does not affect the general trends that we obtain, i.e. our main conclusions, we opted to perform the X-ray spectral fits and UV de-reddening for our final analysis using $N_{\mathrm{H}}=0.36\times 10^{22}~\mathrm{cm}^{-2}$.

The entire model was statistically acceptable with a $\chi^2_\nu < 1.3$ for each fit (for 100--750 dof). We obtained typical values of the photon index in the 1.3--1.8 range and of temperature of the blackbody within 0.4--0.9 keV. We determined the unabsorbed X-ray fluxes (in both the 0.5-10 keV and 2-10 keV range) using the \textsc{cflux} convolution model. 

\subsection{UVOT data reduction and analysis}\label{subsec:uvot}
All the UVOT observations were taken in image mode with one or more filters (see Table~\ref{table: outb}). We calculated the source flux densities with the \textsc{uvotsource} tool, which performs aperture photometry on the sky images. We selected a circular region of radius 5\arcsec\ for the source, and a circular source-free region with a radius of 15\arcsec\ for the background correction. Flux values lower than the 3$\sigma$ limiting flux density in each observation were not included in our further analysis. 
The \source\  optical counterpart in quiescence ($\textit{V}=21.6$ mag ) is contaminated by an interloper star ($\textit{V}=19.4$ mag ) only at 0\arcsec.48 from the source \citep{chevalier1999,hynes2012}. To avoid possible contamination during the outbursts we followed the approach of \citet{meshcheryakov2018_SED_evolution} and subtracted their reported average fluxes for all the UVOT filters. The flux levels were  determined from observations taken when \source\ was in the quiescent state \citep[in 2012 and the pre-outburst period in 2013; see section 3.3 of][]{meshcheryakov2018_SED_evolution}.

We corrected the UV fluxes for the Galactic extinction using the $R_V$-dependent Galactic extinction curve of \citet{fitzpatrick1999} in each filter. We estimated the color excess $E(B-V)$ coefficient in the direction of \source\ using an estimate of $N_{\mathrm{H}}/A_V$ from \citet{predehl1995} for the usual extinction law with parameter $R_V= A_V/ E(B-V)=3.1$. We obtained $E(B-V)= 0.65$~mag with $N_{\mathrm{H}}=0.36 \times 10^{22}~\mathrm{cm}^{-2}$, which agrees with the color excess coefficient estimated from the recalibrated Galaxy extinction maps \citep[][]{schlafly2011}.


\begin{figure*}
\includegraphics[width=0.75\textwidth]{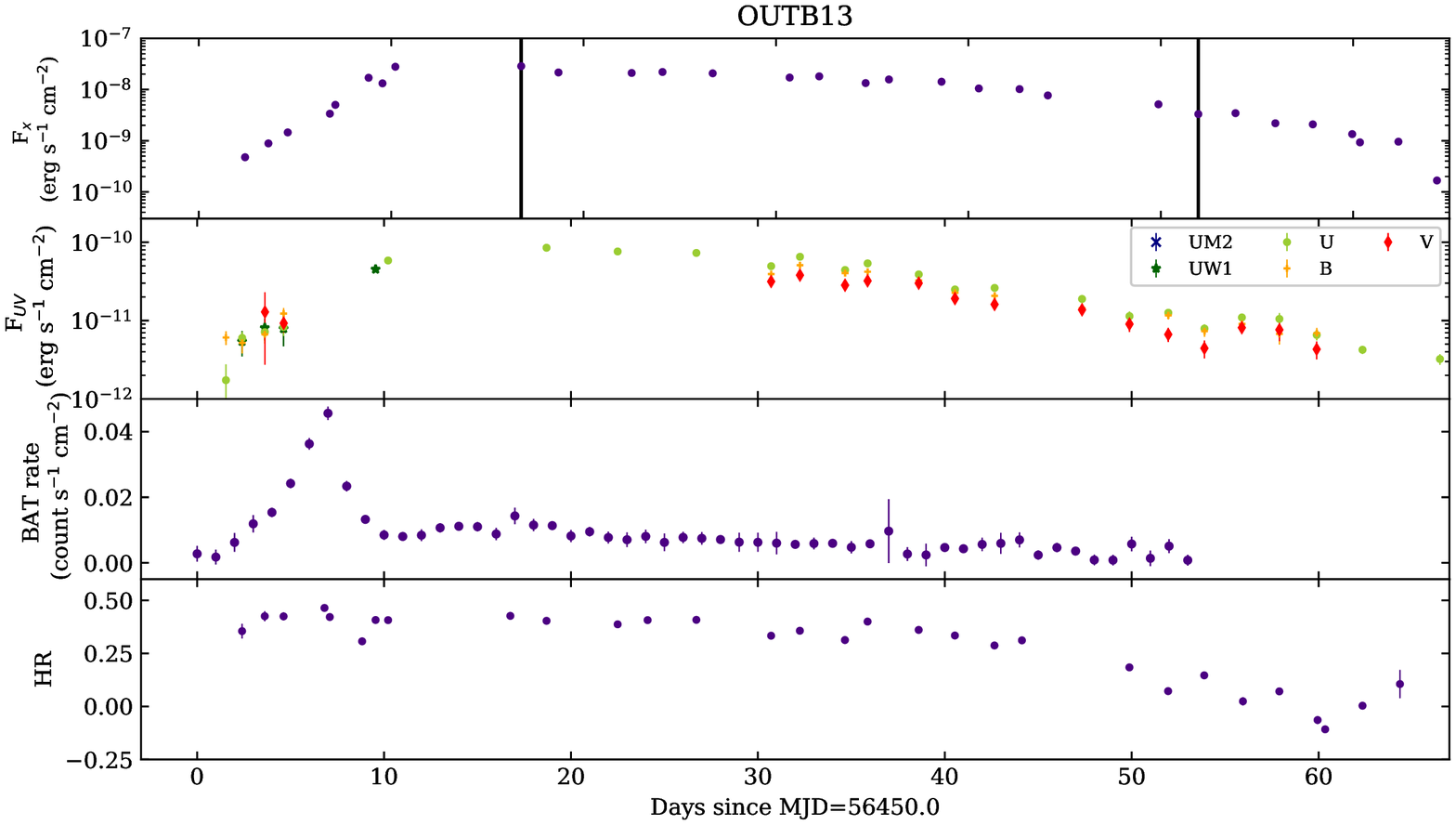}
\includegraphics[width=0.75\textwidth]{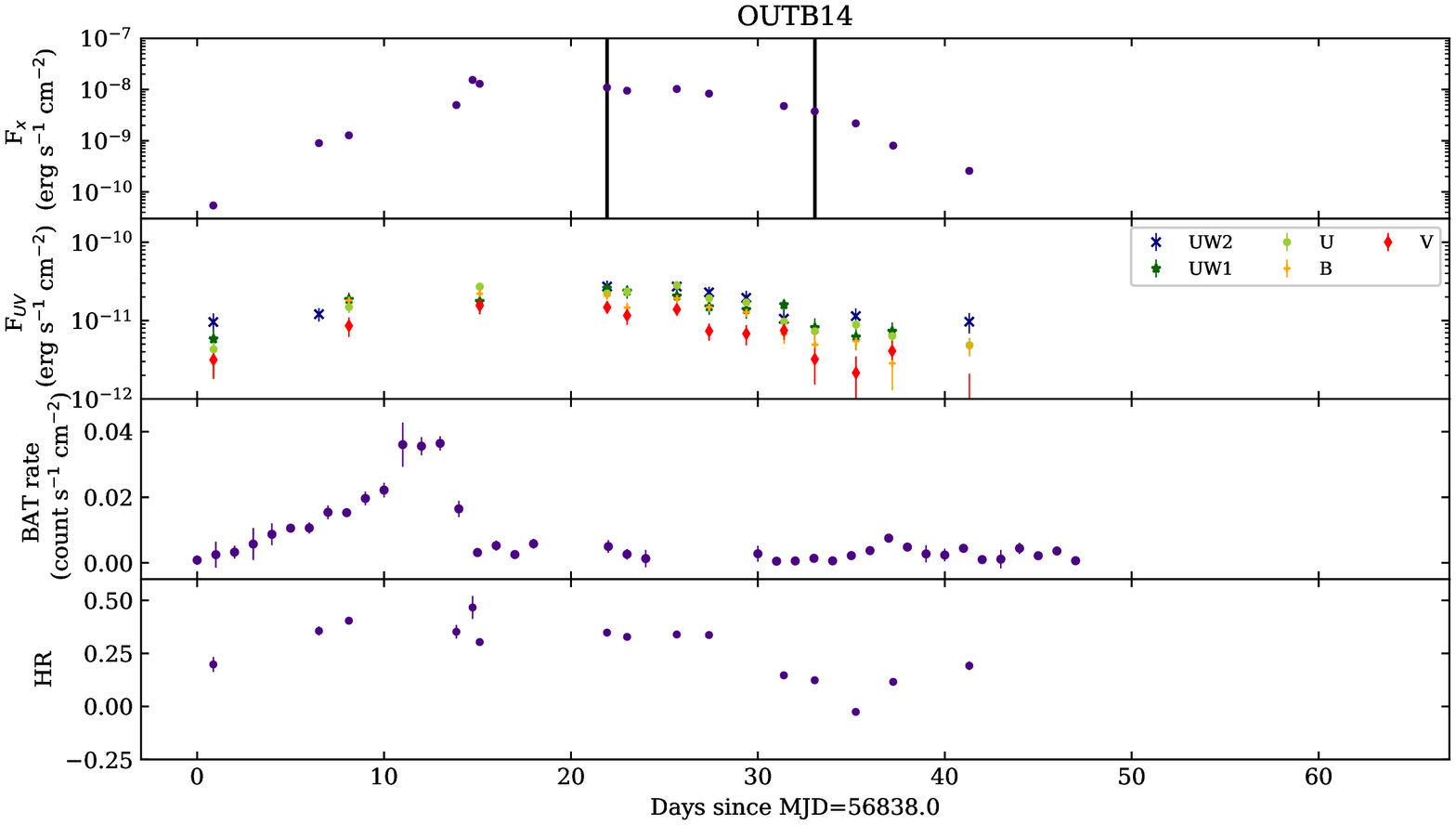}
\includegraphics[width=0.75\textwidth]{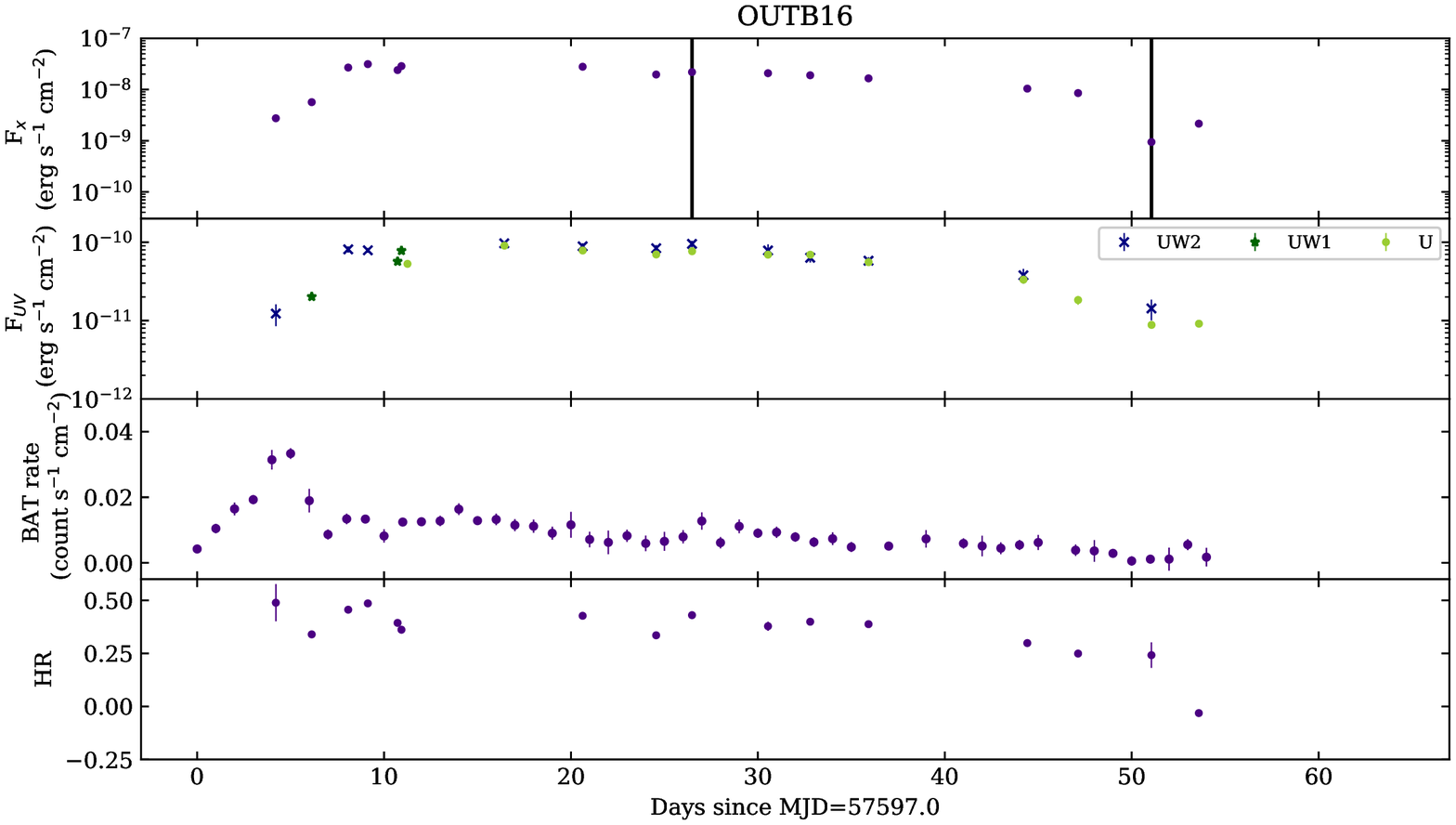}
\caption{Evolution of the 2013 (top), 2014 (middle) and 2016 (bottom) outbursts of \source\ as observed with \swift's XRT and UVOT. For each outburst we show, from top to bottom, the X-ray flux, UV/optical flux, BAT (15-50 keV) rate and 2.5--10/0.5--2.5 keV hardness ratio. The time origin is the first BAT detection (i.e., non zero rate value) for each outburst. Vertical lines in the X-ray light curves mark the limits of the decay before the correlations change (see Figure \ref{fig:correlations}). The plots for each parameter are shown with the same axes limits for each outburst, to allow for a direct comparison. 
}
\label{fig:outbursts}
\end{figure*}

\section{Results}
\subsection{Light curves and spectral evolution}\label{spectral}

The X-ray (0.5--10 keV) and UV/optical light curves for the three outbursts of \source\ are shown in Figure \ref{fig:outbursts}. From these light curves (first two subplots of each outburst) it can be seen that the 2013 and 2016 outbursts had very similar duration and maximum flux, belonging to the long and bright class of FRED, whereas the 2014 outburst was fainter and shorter \citep[see also][]{waterhouse2016,ootes2018}.

To have some sense of the X-ray spectral evolution along the outbursts, we include the daily-averaged 15--50 keV light curves obtained from the \swift/BAT transient monitor project \citep{krimm2013} and the changes in the hardness ratio (HR) in Figure~\ref{fig:outbursts}. In this work, we define HR as the ratio between the \swift\ X-ray flux in the  2.5--10 keV (hard) and 0.5--2.5 keV (soft) bands. We see that the BAT 15--50 keV peak occurs during the rise of the three outbursts, after which there is a drop of the BAT flux that we can identify with the source entering the soft state.

\begin{figure}
\subfigure{\includegraphics[width=\columnwidth]{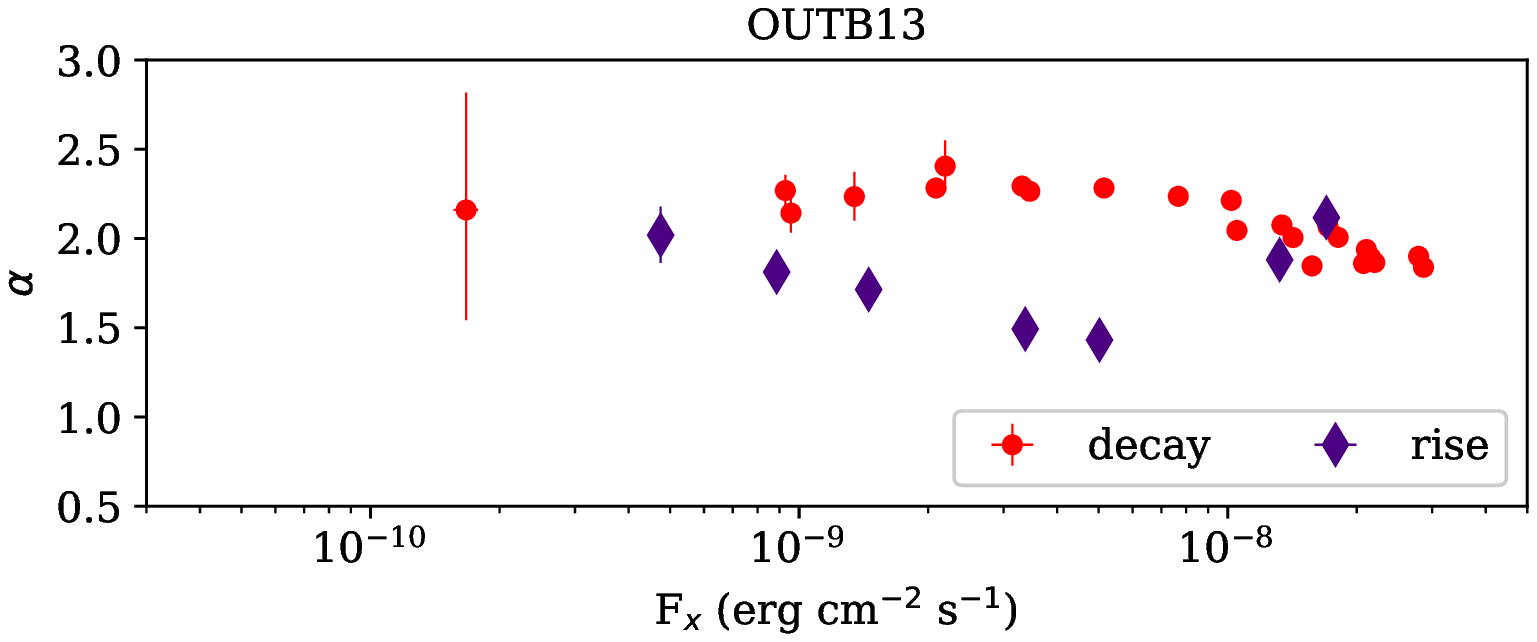}}
\subfigure{\includegraphics[width=\columnwidth]{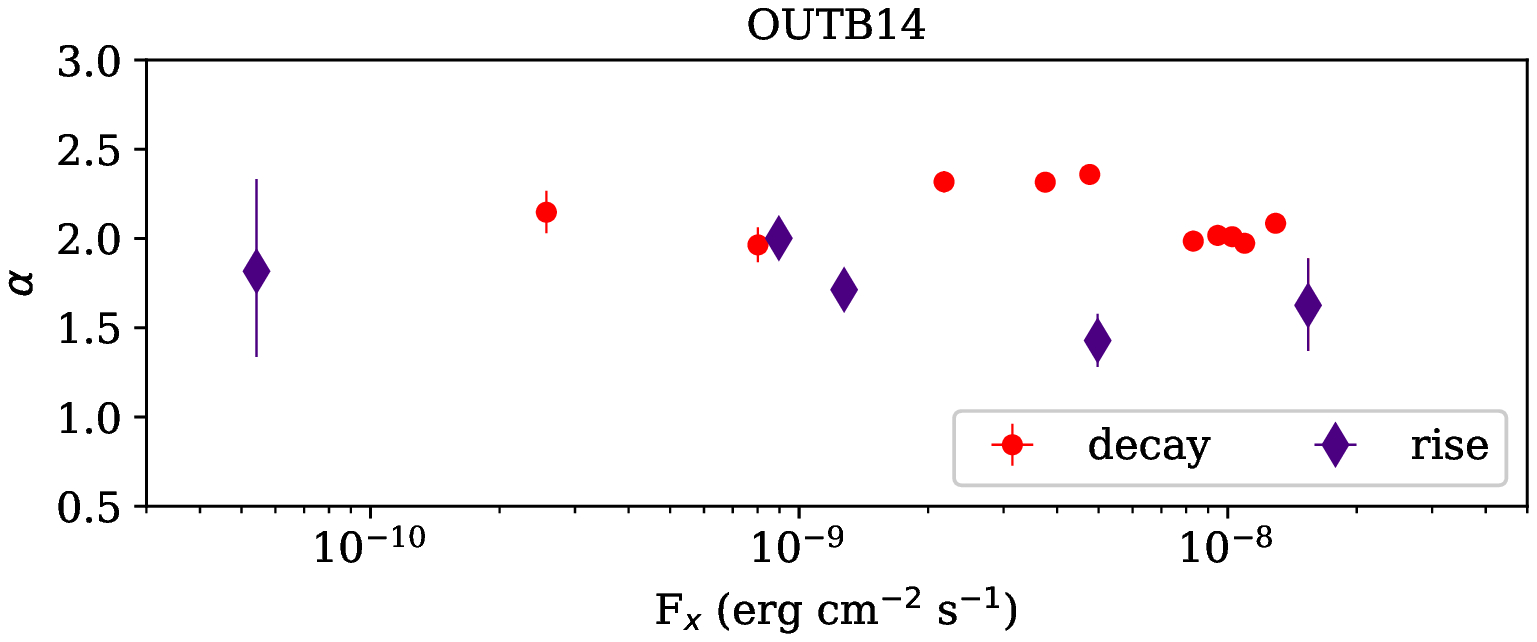}}
\caption{Evolution of the power-law index as a function of X-ray flux during the rise and decay of OUTB13 and OUTB14. At a same flux level, the photon index is smaller during the rise than during the decay, following a hysteresis pattern. }
\label{fig:photonindex}
\end{figure}

In Figure~\ref{fig:photonindex} we plot the photon index $\alpha$ as a function of the X-ray flux, where we highlight the difference between the rise and the decay. To take into account just the variation in the hard X-rays, we plot $\alpha$ obtained from the model \textsc{tbabs*powerlaw} in the range 2.5--10 keV. We only show this for OUTB13 and OUTB14, since OUTB16 had only few observations taken during the rise (see Figure~\ref{fig:outbursts}). It can be seen that at similar X-ray fluxes, the photon index is smaller (i.e. the spectrum is harder) during the rise of the outburst than during the decay, which is likely linked to the peak in the BAT 15--50 keV light curve. Such hysteresis behaviour \citep[][]{miyamoto1995} was noted for \source\ before \citep[][]{maccarone2003}, and is commonly seen in LMXBs, both for BHs \citep[e.g.][]{dunn2011} and NSs \citep[e.g.][]{munozdarias2014}.
The decrease of the HR together with the increase of the photon index indicates that the softening seen along the decays of the outbursts is due to a fall of the hard 2.5--10 keV X-rays, and not to an increase of the thermal emission.

\subsection{The UV/optical and X-ray correlation}\label{subsec:correlation}

\subsubsection{Global properties}\label{subsubsec:global}
The simultaneous observations of \source\ in both X-ray and UV/optical wavelengths allow us to investigate the connection between the UV/optical and X-ray emission along the different outbursts. Figure~\ref{fig:correlations} plots the optical fluxes taken in the U filter against the 2--10 keV X-ray fluxes. Similar plots for other UVOT filters can be found in the Appendix \ref{sec: appendix}. We notice that during the decay of the outbursts the UV/optical and X-ray fluxes seem to follow a correlation. However, the data of the 2013 and 2014 outbursts, for which we have coverage down to lower X-ray fluxes than for the 2016 outburst, suggests that the slope of this correlation changes at F$_{X}\approx (2-4) \times  10^{-9}$~erg~s$^{-1}$~cm$^{-2}$. As we discuss in Section~\ref{subsec:hysteresis}, we speculate that this is because the UV/optical emission becomes dominated by a different process at this time. In Figure \ref{fig:outbursts} we have included vertical lines in the X-ray light curves that mark the limits of the steep decay where the slopes follow an obvious and single trend. We note that these ranges coincide with the spectral softening mentioned in Section \ref{spectral}. 

We proceeded by characterising the correlation between the UV and X-ray fluxes. In doing so, we focussed on observations with X-ray fluxes of $> 10^{-9}~\mathrm{erg~cm}^{-2}~\mathrm{s}^{-1}$, i.e. before the correlations appear to change in Figure~\ref{fig:correlations}. We also computed the correlations during the rise separately when more than two observations were available. On the other hand, in Figure \ref{fig:correlations} we can observe that for OUTB14, the last points of the decay have the same correlation as those of the rise. This is likely linked to the spectral state, as the first and last points of the outburst belong to a hardening phase of the source (see Figure \ref{fig:outbursts}). However, as can be seen in Appendix~\ref{sec: appendix}, we only find this result for the $U$ filter and not for the others, so we chose not to include these last points in our further analysis.  

We fitted the X-ray and UV/optical fluxes assuming a power-law correlation $F_{OUV} \propto F_{X}^{\beta}$, and calculated the correlation slopes $\beta$. To evaluate the significance of the correlations, we calculated the Spearman coefficient ($\rho$), which varies between -1 and 1 with 0 implying no correlation, and the p-value, which roughly indicates the probability of an uncorrelated system producing data sets that have this Spearman correlation. A Spearman correlation coefficient $\rho = +1$ implies an exact monotonic positive relationship. 

We performed our correlation fits for the 2--10 keV X-ray energy range to be able to compare our results with values reported in the literature for the NIR/optical and X-ray correlations of other LMXBs  \citep[e.g.][see also Section~\ref{subsubsec:other}]{russell06}. The results are presented in Table \ref{table:correlations}. We also performed the same analysis using the X-ray fluxes in the 0.5--10 keV band; these results can be found in Appendix~\ref{sec: appendixB}. Although the 0.5--10 and 2--10 keV results are consistent within the errors, during the decay the 0.5--10 keV correlations tend to be steeper, possibly as a result of the spectral softening (i.e., the later the observation the softer it is, so that the 0.5--10 keV flux increases more steeply than the 2--10 keV flux, hence yielding a higher value of $\beta$).

According to our analysis, for each of the data sets the UV/optical and X-ray fluxes are positively correlated during the outburst decay, with a significant $\rho > 0.83$. Comparing the results obtained for the decay of the three different outbursts, we note that the values are formally consistent within the $1\sigma$ errors, which suggests the same emission process for the UV/optical flux in the three outbursts.

\subsubsection{Comparison between the outburst rise and decay}\label{subsubsec:risedecay}
As LMXB outbursts typically show a fast rise and a slower decay, the number of observations sampling the rise is typically much smaller than those obtained during the decay. When a comparison is possible, we observe that the obtained correlation indices tend to be higher during the decay of the outburst than during the rise. 

In Table~\ref{table:correlations}, we highlight in bold the results with $\rho > 0.90$ that suggest a different slope in the rise and decay phases, taking into account the errors. These results suggest that apart from X-ray spectral hysteresis (see Section~\ref{spectral}), we can also see a hysteresis effect in the UV/optical emission. We further discuss this in Section~\ref{subsec:hysteresis}.

\subsubsection{Comparison between the three different outbursts}\label{subsubsec:threeoutbursts}
\begin{figure}
\includegraphics[width=1\columnwidth]{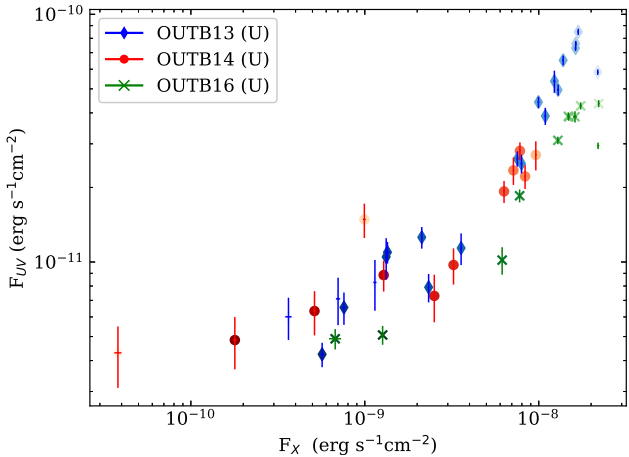}
\caption{Comparison of evolution of the flux in the $U$ filter and 2--10 keV X-ray flux for the three different outbursts.}
\label{fig:all}
\end{figure}

In Figure \ref{fig:all} we directly compare the connection between the optical and X-ray emission for the three different outburst of \source. For this purpose we use the $U$ filter, since it is the only UVOT filter for which we had a reasonably number of data points during all three outbursts. We observe that OUTB14 is much less luminous than OUTB13 and OUTB16 in X-rays, and that the UV/optical emission is comparable between the three outbursts at a given X-ray flux. Moreover, in spite of the similar correlation slopes and 2--10 keV fluxes between OUTB13 and OUTB16, the UV/optical flux at the peak of OUTB13 is considerably brighter than at the peak of OUTB16. We discuss the possible explanations in Section~\ref{subsec:origin}. 

\begin{figure}
\begin{center}
\subfigure{\includegraphics[width=1\columnwidth]{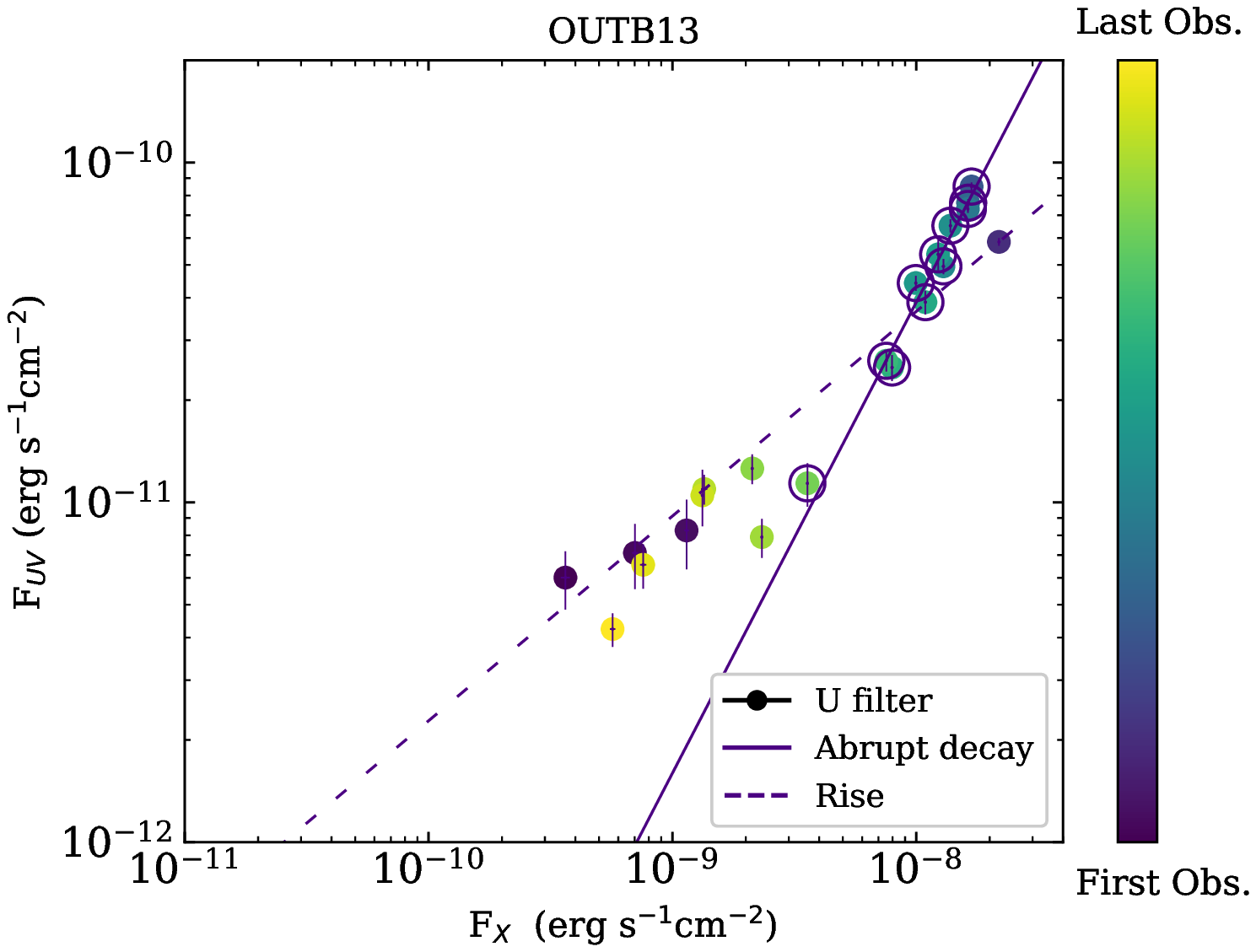}}
\subfigure{\includegraphics[width=1\columnwidth]{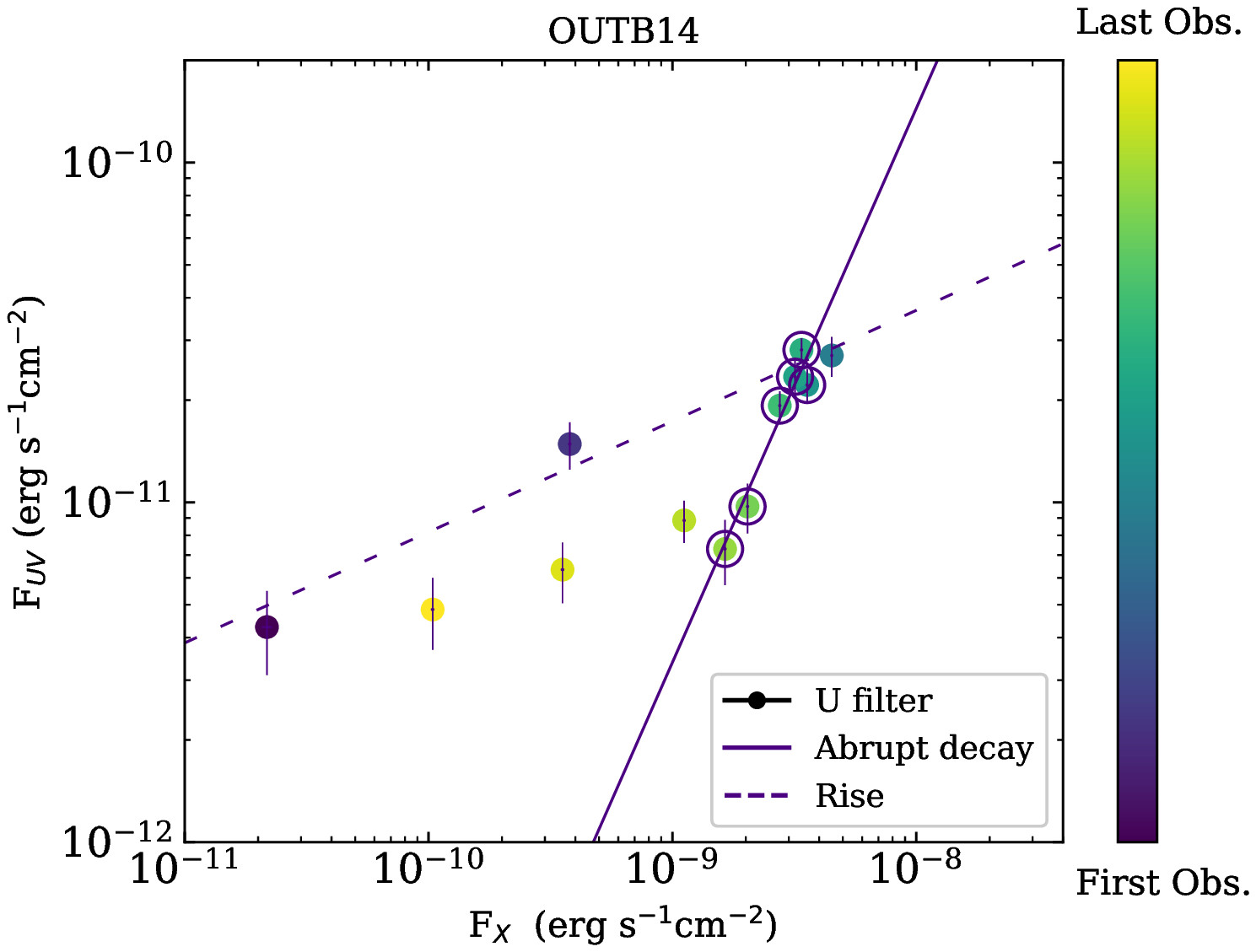}}
\subfigure{\includegraphics[width=1\columnwidth]{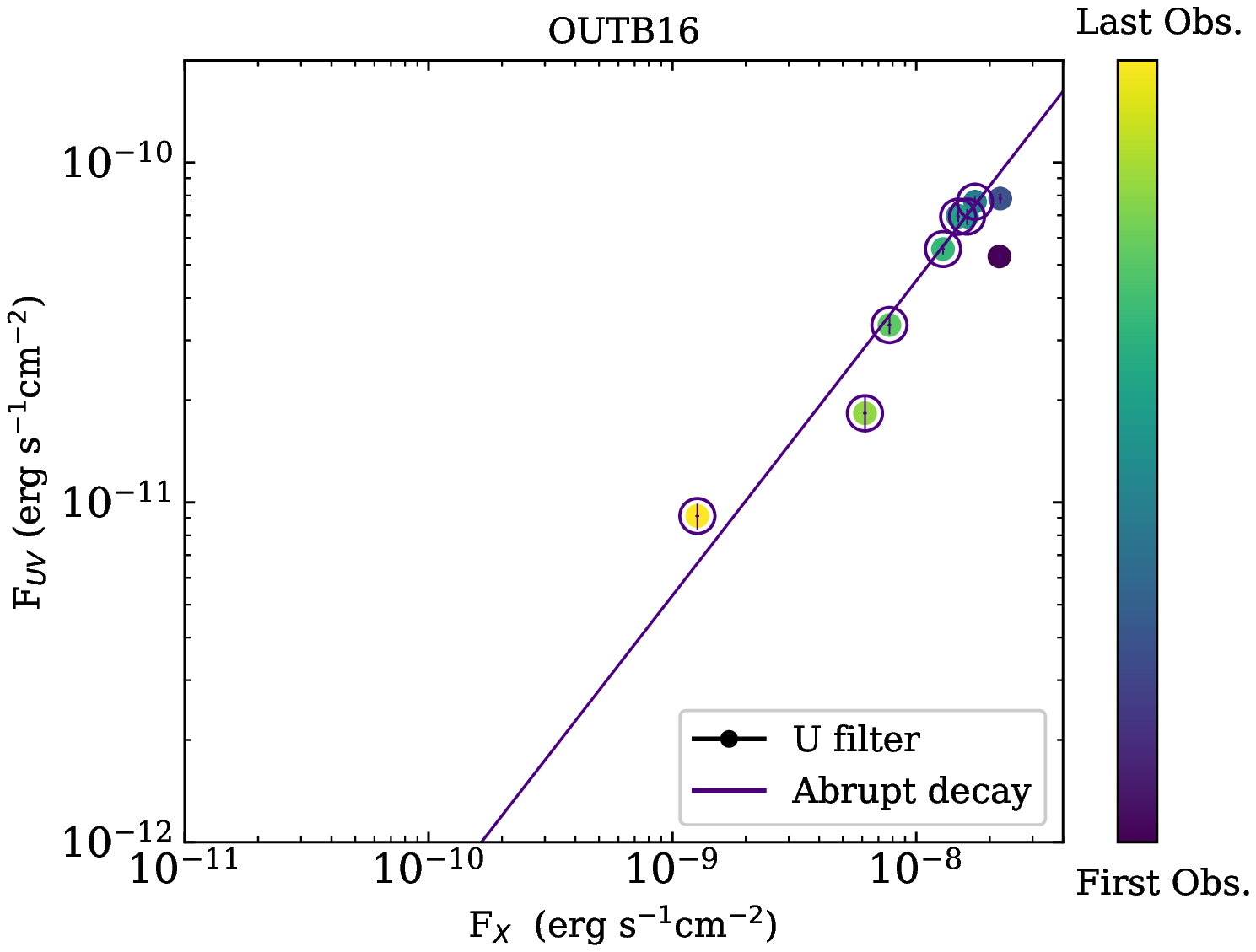}}
\caption{Evolution of the flux in the $U$ filter as a function of the 2--10 keV X-ray flux for the 2013 (top), 2014 (middle) and 2016 (bottom) outbursts. The colour of the markers reflects how far along the outburst the observations were taken, with a darker colour implying an earlier time. The best fits for the rise and the decay are shown. The markers used to fit the decay of the outbursts are circled in the plots (see text for details).}
\label{fig:correlations}
\end{center}
\end{figure}

\begin{table}
\caption{Results of the UV/optical and the 2-10 keV X-ray correlations.} 

\resizebox{0.5\textwidth}{!}{%
\begin{threeparttable}
\begin{tabular}{ccccccc}

\hline\hline
&UVOT filters & \multicolumn{2}{c}{Rise} &\multicolumn{2}{c}{Decay}\\
  & (\#obs rise$/$decay)\tnote{a} & $\beta \pm \Delta \beta $\tnote{b} & $\rho$ (p-value)& $\beta \pm \Delta \beta $\tnote{b} & $\rho$ (p-value)\\

\hline 
\multirow{5}{*}{OUTB13}&UM2 (2/9)&--\tnote{c}& --&1.07$\pm$0.19&0.95 (10$^{-5}$)\\
    &UW1 (4/--)&0.67$\pm$0.05&0.8(0.2)&--&--\\
    &\textbf{U (4/11)}&\textbf{0.60$\pm$0.03}&\textbf{ 1.0 (0.0)}&\textbf{1.39$\pm$0.09}&\textbf{0.97 (10$^{-7}$)}\\
    &B (3/8)&0.8$\pm$0.3& 1.0 (0.0)&1.26$\pm$0.20&0.88 (10$^{-3}$)\\
    &V (2/8)&--& --&1.14$\pm$0.13&0.98 (10$^{-5}$)\\
    \hline 
\multirow{5}{*}{OUTB14}&UW2 (2/4)&--& --&1.00$\pm$0.11&1.0 (0.0)\\
   &UW1 (3/6)&0.20$\pm$0.11&0.5 (0.67)&0.68$\pm$0.23&0.89 (10$^{-2}$)\\
    &\textbf{U (3/6)}&\textbf{0.32$\pm$0.04}&\textbf{1.0 (0.0)}&\textbf{1.04$\pm$0.17}&\textbf{0.83 (10$^{-2}$)}\\
    &\textbf{B (3/6)}&\textbf{0.27$\pm$0.15}&\textbf{1.0 (0.0)}&\textbf{1.18$\pm$0.12}&\textbf{0.94(10$^{-3}$)}\\
    &\textbf{V (3/6)}&\textbf{0.29$\pm$0.01}&\textbf{1.0 (0.0)}&\textbf{1.0$\pm$0.3}&\textbf{0.94 (10$^{-3}$)}\\
    \hline
\multirow{3}{*}{OUTB16}&UW2 (3/7)&0.75$\pm$0.17&0.5 (0.67)&1.2$\pm$0.3&0.83 (10$^{-2}$)\\
    &UW1 (3/--)&0.73$\pm$0.11&1.0 (0.0)&--&--\\
    &U (1/7)&--& --&1.20$\pm$0.11&0.89 (10$^{-3}$)\\
\hline
\end{tabular}
\begin{tablenotes}
  \item[a] The number of observations obtained during the rise/decay are given in parenthesis.
  \item[b] Errors reflect 1$\sigma$ confidence intervals.
  \item[c] We only attempted to fit correlations when 3 or more observations were available.
  \item[] The values indicated in bold highlight the filters/outbursts where the rise (hard to-soft state) gives a significantly different correlation than the decay (soft state).
  \end{tablenotes}
\label{table:correlations}
\end{threeparttable}
}

\end{table}

\section{Discussion}\label{sec:discuss}

\subsection{The UV/optical and X-ray flux correlation}\label{subsec:origin}
In LMXBs, different UV/optical emission processes are expected to be connected in different ways (i.e. yielding different correlations) with the X-ray flux. \citet{russell06} studied an ensemble of NS and BH LMXBs to estimate the contributions of various processes to the NIR/optical emission. These authors found a global correlation for their sample of 8 NS systems in the hard state of $L_{OIR} \propto L_{X}^{0.63\pm 0.04}$, which holds over 7 orders in magnitude in $L_X$ ($10^{31}$< $L_X$<$10^{38}$ erg s$^{-1}$). According to \citet{russell06}, the observed correlation can be explained as X-ray reprocessing being the dominant NIR/optical emission process, possibly with contributions of the viscously heated disc and, at high luminosity, from a jet. 
Here, we compare our results obtained for the \textit{U}, \textit{B} and \textit{V} filters with the $\beta$ values in the optical waveband that are expected from the theoretical models, in an attempt to investigate which processes are more likely to dominate the UV/optical emission.

\subsubsection{Outburst rise: from hard to soft state}\label{subsubsec:rise}
For OUTB13 and OUTB14 we see that the correlation indices are systematically smaller during the rise, which suggests that different UV/optical emission mechanisms may be dominating at different spectral states. We find values of $0.6<\beta_{UBV}<1.1$ for the rise of OUTB13, and $0.2<\beta_{UBV}<0.4$ for OUTB14. Although the errors are large and we have more limited UVOT coverage for OUTB16, our results do suggest that the correlation index between the UV/optical and X-ray fluxes was also lower during the rise than during the decay of OUTB16. 

The values that we obtain for the rise of the OUTB13 appear to be broadly consistent with those of \citet{russell06}, and what is theoretically expected for X-ray reprocessing  \citep[$\beta_{rep} \simeq 0.5$;][]{vanparadijs1994}. 
Therefore, this could plausibly be the dominant UV/optical emission process during the rise of OUTB13. This is likely also true for the 2016 outburst, for which we obtain similar correlation indices.

On the other hand, the $\beta$ values obtained for the 2014 outburst are much lower than seen for the 2013 outburst, and expected for X-ray reprocessing. Instead, the results for the OUTB14 rise are consistent with emission from the viscously-heated disc dominating the UV/optical emission, for which $0.30<\beta_{disc}<0.60$ is expected for NS LMXBs \citep[][]{frank2002}. 

However, we have to be cautious in drawing strong conclusions about these results, as there is a change of the spectral state along the rise of the outbursts. According to Figure~\ref{fig:outbursts}, the last points of the rises belong to the soft state, after the drop of the 15--50 keV BAT flux. This spectral state change would likely lead to a different UV/optical emission, due to the drop of 15--50 keV flux and the change of morphology of the disc. Different results are found during the decay of the outbursts, where the source is in a soft state.


\subsubsection{Outburst decay: soft state}\label{subsubsec:decay}
Along the decay, the UV/optical correlations found in the three outbursts have values in the range $0.7<\beta_{UBV}<1.5$. Although the errors of our individual fits are large, it is clear that our obtained slopes are systematically higher than those obtained during the rise (see Section~\ref{subsubsec:rise}). We discuss this difference in more detail in Section~\ref{subsec:hysteresis}. Moreover, our $\beta$ values for the outburst decays are also higher than the results obtained by \citet{russell06}, and the correlation expected from X-ray reprocessing  \citep[][]{vanparadijs1994}. The fact that we find higher correlation coefficients suggests that X-ray reprocessing alone likely cannot explain our observed UV/optical emission in the soft state. On the other hand, we note that a steeper correlation could be expected if the UV/optical emission comes also from the reprocessing of more energetic X-rays (> 10 keV). 

We note that the daily-averaged BAT count rate and the UVOT $U$ flux by eye appear to be strongly correlated during the decay of OUTB13 and OUTB16. However, the X-ray spectrum of \source\ is soft and there is more hard ($>2$~keV) flux in the XRT band than in the BAT band. Therefore, we do not expect a big influence of the BAT flux in the correlations found during the decay of the outbursts. This likely implies then that multiple mechanisms are contributing to the UV/optical light observed from \source. 

Looking at the results of OUTB13, we see that the slopes of the correlation tend to increase towards shorter wavelengths, which is expected for UV/optical emission from irradiation but also from a viscously-heated disc \citep{frank2002}. This suggests that in the decay of the outburst where the source is in a soft state, the direct emission from the disc could also contribute to the UV/optical emission. 

We note that most of the correlation indices that we obtain during the decay are consistent within the errors for the three outbursts, suggesting that the UV/optical emission of \source\ is dominated by the same processes in the class of bright and long outbursts, such as the 2013 and 2016 ones, and the class of fainter and shorter outbursts, to which the 2014 one belongs. This is in agreement with the work of \citet{maitra2008}, who found a similar optical/X-ray flux behaviour for several FRED outbursts. However, in Figure~\ref{fig:all} we saw that the UV/optical fluxes at the beginning of the OUTB16 decay are lower than those for OUTB13. The explanation cannot be found in the X-ray reprocessing origin, as the 2--10 keV and 15--50 keV fluxes are similar for both outbursts at these points. Other mechanisms must be leading to this flux difference. In the case of the viscously heated disc, a difference in the heating mechanism would lead to a increase or decrease of the UV/optical flux. Interestingly, the viscous time scale from the outer disc in LMXBs is of the order of days-weeks \citep[e.g.][]{frank2002}. If the heating mechanism is related to the BAT 15--50 keV peak then the lower UV flux of OUTB16 with respect to OUTB13 could be related to the lower 15--50 keV BAT rate at the peak. 

It is worth noting that \citet{migliari2006} studied the relation between the radio (jet) and X-ray fluxes of a sample of NS LMXBs and suggested that for some sources these may be correlated as $\beta_{jet}= 1.4$. If the jet emission would happen to be flat from the radio to the optical waveband, our obtained correlation indices for the UV/optical emission could potentially suggest a jet contribution. Although \source\ is known to display radio jet emission during its outbursts, including the ones studied in this work \citep[e.g.][]{tudose2009,millerjones2013_aqlx1,diaztrigo2018,gusinskaia2019}, we do not deem this scenario very likely. Firstly, it appears that the radio/X-ray correlation in \source\ has a much lower $\beta_{jet}$ value \citep{tudose2009,migliari2006,tetarenko2018,gusinskaia2019}. Moreover, a recent radio/X-ray study, using the largest sample of NS LMXBs to date, also found a lower correlation index for the population as a whole \citep[$\beta_{jet}= 0.44$;][]{gallo2018}. Finally, detailed multi-wavelength studies of \source\ and other systems suggest that the jet of NS LMXBs is not likely to contribute significantly to the emission at NIR/UV/optical wavelengths, either due to NSs having weaker jets than BHs, or due to the jet breaking at a lower frequency \citep[e.g.][]{migliari2006_4u0614,russell06,maitra2008}.  


Apart from the disc and the jet, a hot flow or emission from the NS magnetosphere could possibly produce UV emission \citep[e.g.][]{kargaltsev2007,veledina2011}. However, for the latter scenario we may expect to see signs of a dynamically important magnetic field, which is not obvious for \source.\footnote{Coherent X-ray pulsations were detected from \source\ only during a very short instance of time; the source otherwise behaves as any non-pulsating NS LMXB \citep[][]{casella2008}. On the other hand, reflection studies in both the hard and the soft state suggest that the inner disc could potentially be truncated by the neutron star magnetic field \citep[][]{king2016,ludlam2017_aqlx1}.} Emission from some form of hot flow, on the other hand, could be plausible \citep[e.g.][]{esin1997,veledina2011,veledina2013}. One might expect that the hot flow grows as the accretion rate decreases and hence that the ratio of UV/optical over X-ray flux increases with decreasing X-ray luminosity. Perhaps this can account for the flattening of the flux correlation that we see for OUTB13 and OUTB14 when the X-ray flux decreases below $\simeq 10^{-9}~\mathrm{erg~cm}^{-2}~\mathrm{s}^{-1}$. This flux level translates into an X-ray luminosity of $\simeq 10^{36}~\mathrm{erg~s}^{-1}$ at the estimated distance of \source\ ($\sim$5~kpc), and roughly corresponds to $\sim$1\% of the Eddington luminosity for a NS. It seems plausible that a hot flow becomes an important UV/optical emission mechanisms at low luminosities as an LMXB transitions towards quiescence \citep[e.g.][]{hynes1999}.

Alternatively, if a large fraction of the in-falling gas is expelled in an outflow, this may also cause an apparent excess UV/optical emission. This is because if the material that generates the UV/optical emission in the outer parts of the disc is not reaching the inner part, this material will not produce X-rays. In LMXBs, it appears that disc winds may remove a substantial amount of gas from the disc \citep[e.g.][]{miller2006_winds,neilsen2009,ponti2012_winds,munozdarias2016}. For \source\ there are no reports of detected disc winds. However, the density of disc winds is concentrated towards the disc plane \citep{higginbottom2017}, so if the inclination of the binary is relatively low, it will be difficult to detect a disc wind, if present.\footnote{In case of \source, both NIR spectroscopy and X-ray reflection studies suggest a relatively low disc inclination of $\lesssim 50^{\circ}$ \citep[][]{ludlam2017_aqlx1,matasanchez2017_aqlx1}, but see \citet{galloway2016}.}


\subsubsection{Comparison with other LMXBs}\label{subsubsec:other}

A similar analysis as we perform here, i.e. quantifying any correlation between the UV/optical and X-ray fluxes using \swift\ data, has been carried out for a number of other BH and NS LMXBs. In Table \ref{table: sources}, we present an overview of the correlation indices obtained for other sources from the literature, where we also list the interpretation given in different studies. Although X-ray reprocessing has been claimed as the main mechanism for the UV/optical emission in LMXBs, studies for individual sources do not agree on a single origin. Moreover, none of the sources has been studied in a soft state nor along the entire outburst. The most similar result was $\beta=1.00^{+0.34}_{-0.14}$ derived by \cite{cackett2013_cenx4} for the NSXB Cen X-4 during its quiescent state. However, they could not make strong conclusions about the form of the correlation as the dynamic range analysed in L$_{x}$ was too small.

\begin{table*}

\caption{Summary of optical/NUV and X-rays correlations of other LMXBs from the literature (studied using \swift\ data).}
\begin{center}

\begin{adjustbox}{max width=\textwidth}
\begin{tabular}{ccccc}
\hline\hline
Type & Source & UVOT filters & $\beta$ & Results/Interpretation \\
\hline 

\multirow{4}{*}{\textbf{BH}} & \multicolumn{1}{c}{Swift J1357.2-0933} & \multicolumn{1}{c}{\textit{UW2,UW1,U,B,V}}& \multicolumn{1}{c}{$0.2 < \beta < 0.4$}& \multicolumn{1}{c}{Viscously heated disc emission \citep[hard state;][]{armas2012}}\\
& \multicolumn{1}{c}{Swift J1910.2-0546} & \multicolumn{1}{c}{\textit{UM2}}& \multicolumn{1}{c}{from $-0.45 < \beta < 0.15$}& \multicolumn{1}{c}{Time lag between X-ray and UV emission \citep[from soft to hard state;][]{degenaar2014}}\\
& \multicolumn{1}{c}{GX 339-4} & \multicolumn{1}{c}{\textit{UW2}}& \multicolumn{1}{c}{0.50}& \multicolumn{1}{c}{Jet emission \citep[from hard to soft state;][]{yan2012}} \\
& \multicolumn{1}{c}{XTE J1817-330}&\multicolumn{1}{c}{\textit{UW1}}& \multicolumn{1}{c}{0.50}& \multicolumn{1}{c}{Reprocessed emission \citep[from soft to hard state;][]{rykoff2007}} \\
                                 
\hline
\multirow{3}{*}{\textbf{NS}} & \multicolumn{1}{c}{Cyg X-2} & \multicolumn{1}{c}{\textit{UW2}}& \multicolumn{1}{c}{uncorrelated}& \multicolumn{1}{c}{Anticorrelation between the NUV and the hard X-ray color and the BAT flux \citep[Z-source;][]{rykoff2010}} \\
& \multicolumn{1}{c}{SAX J1808.4-3658} & \multicolumn{1}{c}{\textit{UW2,UM2,UW1,U,B}}& \multicolumn{1}{c}{$0.15 < \beta < 0.3$}& \multicolumn{1}{c}{Viscously heated disc emission \citep[AMXP, hard state;][]{patruno2017_1808}}\\
& \multicolumn{1}{c}{Cen X-4} & \multicolumn{1}{c}{\textit{UM2,UW1}}& \multicolumn{1}{c}{$1.00^{+0.34}_{−0.14}$}& \multicolumn{1}{c}{No correlation with the optical filters \citep[quiescence;][]{cackett2013_cenx4}} \\

\hline 

\end{tabular}
\end{adjustbox}

\label{table: sources}

\end{center}
\end{table*}

\subsubsection{Assumptions and applicability}
In this work, we have assumed that there should be a power-law relationship between X-rays and UV/optical for a disc irradiation model, and that the failure of a single power-law correlation requires multiple emission processes. However, while other mechanisms such as viscous disc heating may well be playing a role, there are also other factors that can modify the relation between X-rays and UV/optical even in a ``simple'' disc irradiation scenario.

Firstly, since we observe the reprocessed light through a limited bandpass, the steepness of this relation depends on where the observed bandpass falls with respect to the peak of the reprocessed spectrum. The actual value will depend on the filter used and the temperature of the reprocessing material. We  indeed observe some of this effect, as we note in Section~\ref{subsec:correlation}. This can be seen by comparing the correlations we computed in the 2--10 keV band to those obtained when using the broader 0.5--10 keV band (given in Table \ref{table:correlations05}). 

Secondly, there is the basic assumption that the observed X-ray flux traces the central luminosity and that the reprocessed emission varies linearly with the central luminosity. These assumptions could be violated by X-ray spectral changes, particularly dramatic state changes, varying emission geometry, absorption or albedo (e.g. by Compton reflection of the illuminating X-ray flux). This is in fact what we believe is leading to the different correlations observed in the outbursts rise, where \source\ shows a spectral state change (from hard to soft).
In the decay, there is no obvious X-ray spectral state change, albeit geometrical changes in the accretion flow could be violating the applicability of the power-law relation. 
These considerations should be kept in mind.


\subsection{Hysteresis}\label{subsec:hysteresis}
From our X-ray spectral analysis, we recovered the hysteresis behaviour commonly seen in LMXBs \citep[e.g.][]{dunn2010,munozdarias2014}: we found that the X-ray spectrum was systematically softer during the decay of the 2013 and 2014 outbursts than it was during the rise. Such hysteresis cannot be explained by the disc instability model \citep[e.g.][]{hameury2017}, and must have a different physical cause. Proposed explanations include a role of the disc magnetic field \citep[e.g.][]{balbus2008,petrucci2008,begelman2014}, and Lense-Thirring procession of the inner disc \citep[e.g.][]{nixon2014}, but hysteresis remains poorly understood.

In addition to the commonly known X-ray spectral hysteresis, we now also observe a hysteresis pattern in the UV/optical emission of \source; at a given X-ray flux, the observed UV flux is lower during the decay of the outburst than it is during the rise. This indicates that the X-ray spectral behaviour is linked to the UV/optical emission as we have been seeing along this work. We note that \citet{maitra2008} studied \source\ over multiple outbursts of different classes and do not report any hysteresis behaviour in the optical/NIR emission.

\begin{figure*}
\subfigure{\includegraphics[width=0.695\columnwidth]{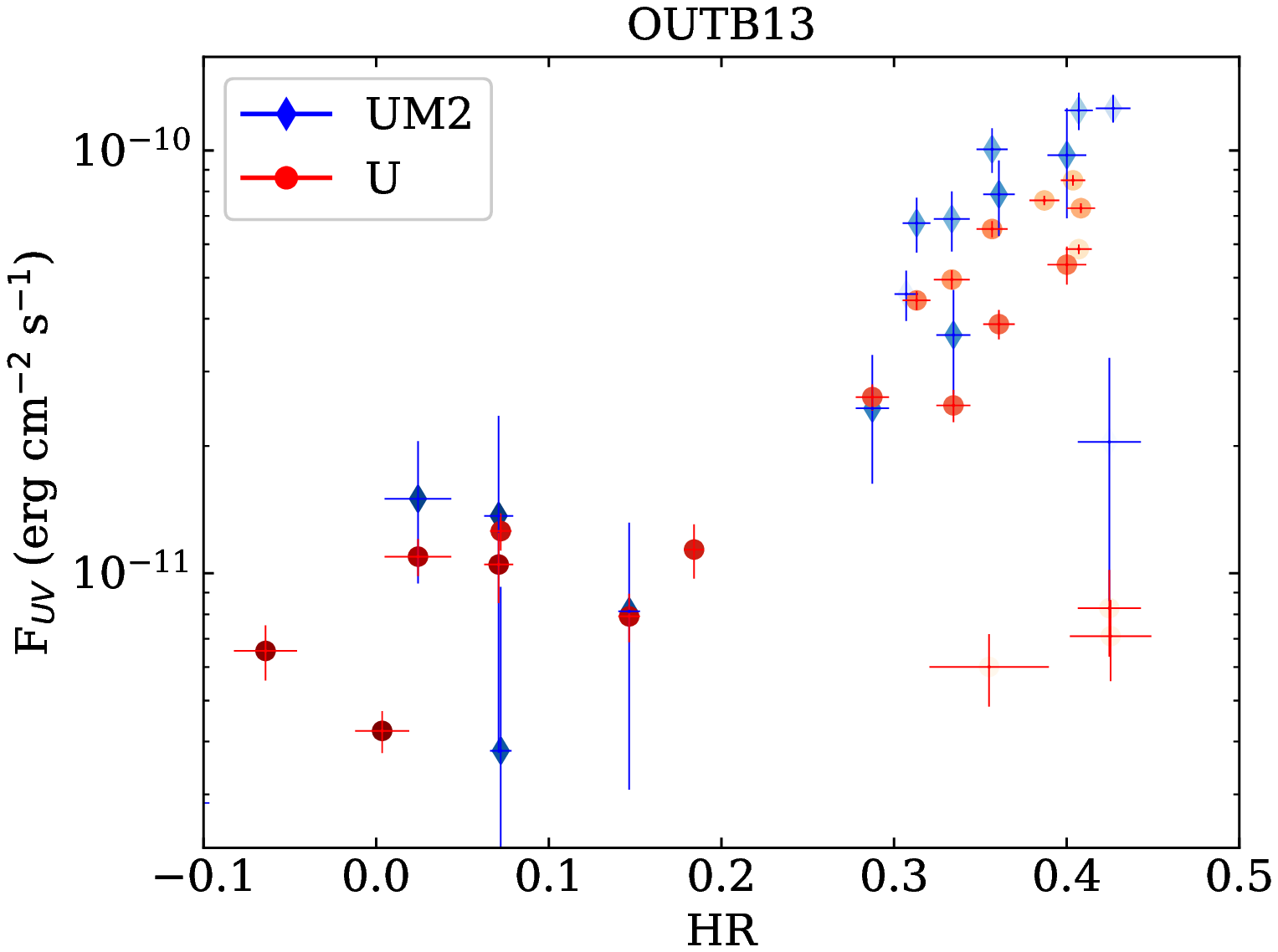}}
\subfigure{\includegraphics[width=0.695\columnwidth]{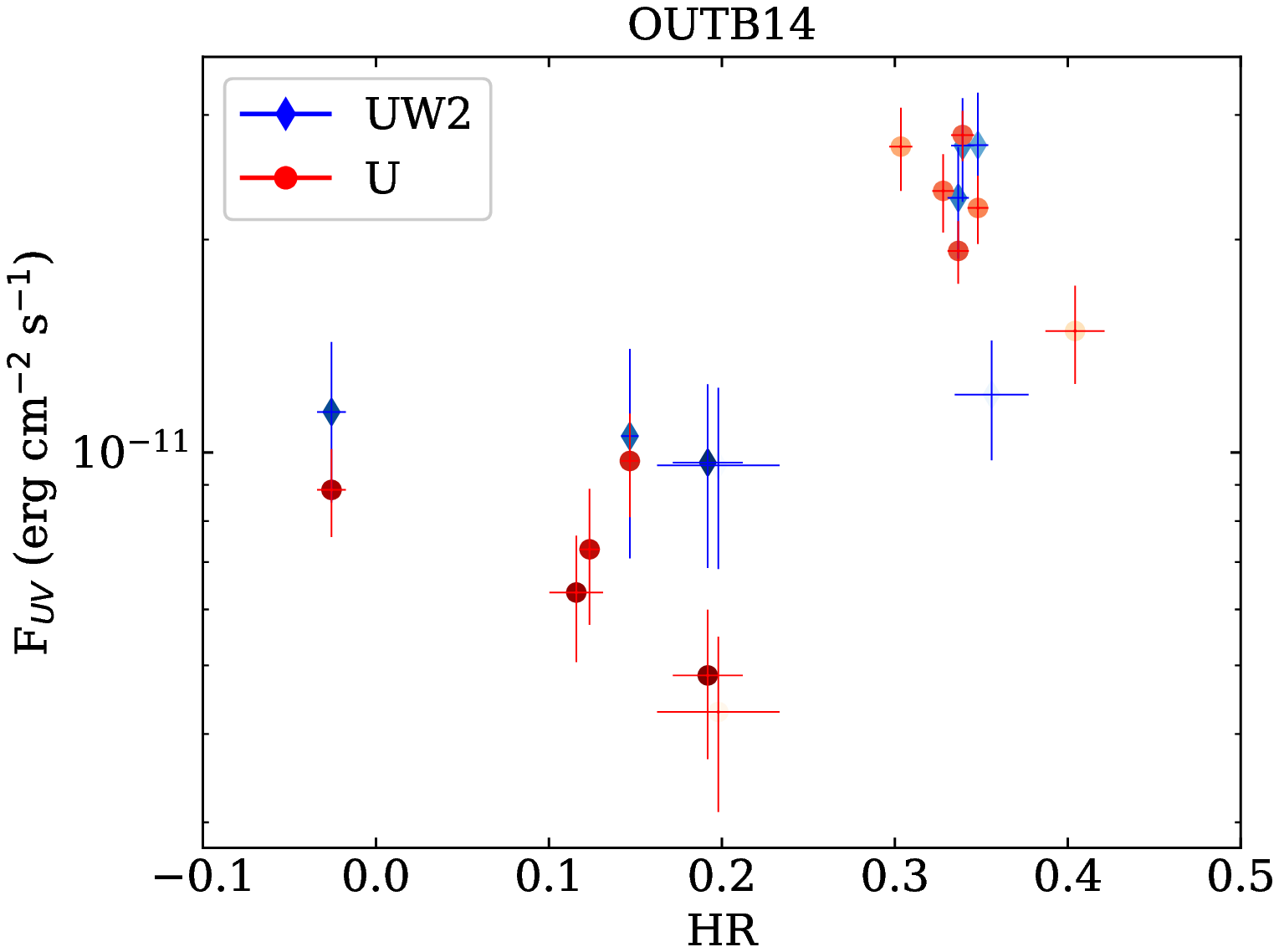}}
\subfigure{\includegraphics[width=0.695\columnwidth]{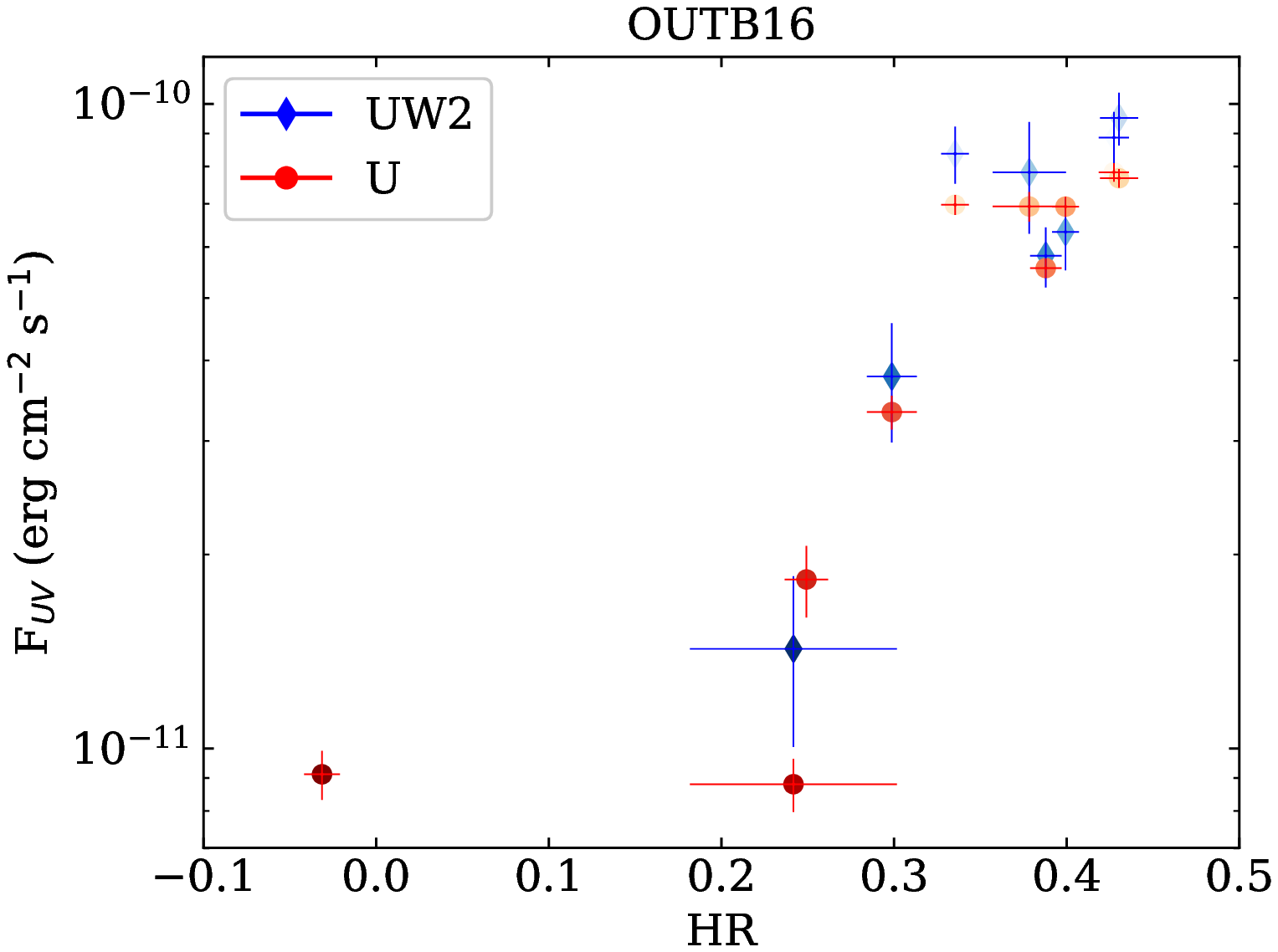}}

\caption{The UV/optical fluxes as a function of X-ray hardness ratio for the three outbursts studied in this work. The brightness of the data points indicates the time along the outburst (with darker points occurring at earlier times).}
\label{fig:hardness}
\end{figure*}

Exploring the correlation between the X-ray HR (see Section~\ref{spectral}) and the UV/optical fluxes provides further clues about the origin of this emission and the observed hysteresis behaviour. In Figure~\ref{fig:hardness}, we show the correlation between the UV/optical fluxes and the X-ray HR for the three outbursts. Looking at the temporal evolution (i.e. the brightness coding) together with the UV/optical and HR during OUTB13 and OUTB14, we see again the hysteresis pattern mentioned above. First, a hard state is observed in the initial rise, which is followed by a softening of the source and a decrease in the UV/optical emission. We see that a hard state does not directly imply a high flux in the UV/optical wavebands, but it is related to the drop of BAT flux and the entering to the soft state as we showed in Section \ref{spectral}. 

The softening of LMXBs during an outburst rise is generally assumed to imply that the thermal accretion disc emission is becoming increasingly  important in the X-ray energy spectrum. Therefore, we can interpret the increase of the UV/optical flux as a direct consequence of a change in the accretion geometry/morphology, either due to  higher UV/optical emission from the viscously-heated disc itself, or to an increase of the hard X-rays reprocessed in the disc. At some point, around HR$\simeq0.2$, the hardness ratio still decreases but the UV/optical emission keeps approximately constant, until the system becomes harder again and F$_{\mathrm{OUV}}$ decreases. 

The hysteresis pattern described above, and shown in Figure~\ref{fig:hardness}, is consistent with the X-ray reprocessing model as the main UV/optical emission mechanism. We speculate that there is an inflection point in the softening phase where the emission from the viscously-heated disc becomes dominant, leading to a different behaviour of the UV/optical emission. This scenario explains satisfactorily the observations of OUTB13 and OUTB14 at later times (Figure \ref{fig:hardness}). For OUTB16, we see the rise phase in the UW2 filter and the point of inflection in the U filter, but we cannot observe the hardening phase due to limited number of observations. In the last hardening phase, the disc would change again its morphology leading to a lower UV/optical emission.
We note that in OUTB14 this decrease seems to be more important for the $U$ filter than for the $UW2$ one. This result favours an origin for the UV emission probed by the $UW2$ filter as being located in the hot inner region of the disc as also suggested in previous works \citep{campana2000,mcclintock2003,hynes2012,cackett2013_cenx4}. It then appears that changes in the morphology of the corona and the inner disc more strongly affect the outer parts of the disc, where  less energetic optical emission is originating.

\subsection{Summary $\&$ conclusions}\label{subsec:conclusions}

In this work, we investigated the connection between the UV/optical and X-ray fluxes of the neutron star LMXB \source, using \swift\ data obtained during its 2013, 2014 and 2016 outbursts. Our main findings are:

\begin{itemize}
    \item We find a strong correlation between the UV/optical and X-ray fluxes during the decay of the three outbursts, where the source is in a soft state. Quantifying this correlation suggests that the 2013 and 2016 outbursts, which both belong to the class of long/bright outbursts of \source, and the fainter/shorter 2014 outburst behaved similar. The values that we obtain for \source\ during the decays are very different to that of a number of other BH and NS LMXBs in the hard state for which similar studies were performed. This is likely due to the different spectral state of the sources from the literature. \\
    
    \item The 2013 and 2014 outbursts had sufficient coverage to investigate the rise and decay separately. This revealed that for both outbursts the UV/optical and X-ray correlation is significantly different during the rise (hard-to-soft state) than during the decay (soft state). This is likely linked to the commonly observed X-ray spectral hysteresis seen in LMXBs, as a change in the disc morphology could lead to a different UV/optical emission either due to higher emission from the viscously-heated disc itself, or to an increase of the hard X-rays reprocessed in the disc. \\ 

    \item The X-ray reprocessing model alone is not likely to account for the correlation indices obtained during the soft state. Thus, we suggest that multiple emission processes are contributing to the observed UV/optical emission of \source, such as the viscously heated disc or a hot flow. We do caution that the limited passband of the observations and model assumptions may give correlation indices that deviate from the theoretical predictions.  \\
    
\end{itemize}

Our study reinforces that for LMXBs with sufficiently low extinction ($\lesssim 5 \times 10^{21}~\mathrm{cm}^{-2}$), UV studies can provide valuable information about the accretion process. Capturing both the hard and soft spectral states of an outburst is particularly valuable in this respect. With its flexibility and multi-wavelength capabilities, \swift\ is a very suitable tool to perform such studies. In order to reach firm conclusions, however, an entire outburst needs to be densely monitored (every few days) from the start till the end, and use a consistent set of UVOT filters (rather than using ``filter of the day"). 

\vspace{-0.2cm}
\section*{Acknowledgements}
ELN gratefully acknowledges support from an "ERASMUS+ for traineeship" grant, and is thankful for the hospitality of the Anton Pannekoek Institute, where most of this work was carried out. ND, ASP, JVHS, JvdE are supported by a Vidi grant from the Netherlands Organization for Scientific research (NWO), awarded to ND. JVHS acknowledges support from a STFC grant ST/R000824/1. This research made use of {\sc astropy}, a community-developed core {\sc python} package for Astronomy \citep{Astropy-Collaboration:2013aa} and {\sc matplotlib} \citep{Hunter:2007aa}. 
This work made use of data supplied by the UK Swift Science Data Centre at the University of Leicester.

\footnotesize{
\bibliographystyle{mnras}
\bibliography{thesis}
}
\appendix
\section{Additional UV/optical and X-ray flux plots}
\label{sec: appendix}

\begin{figure*}
\begin{center}
\subfigure{\includegraphics[width=0.8\columnwidth]{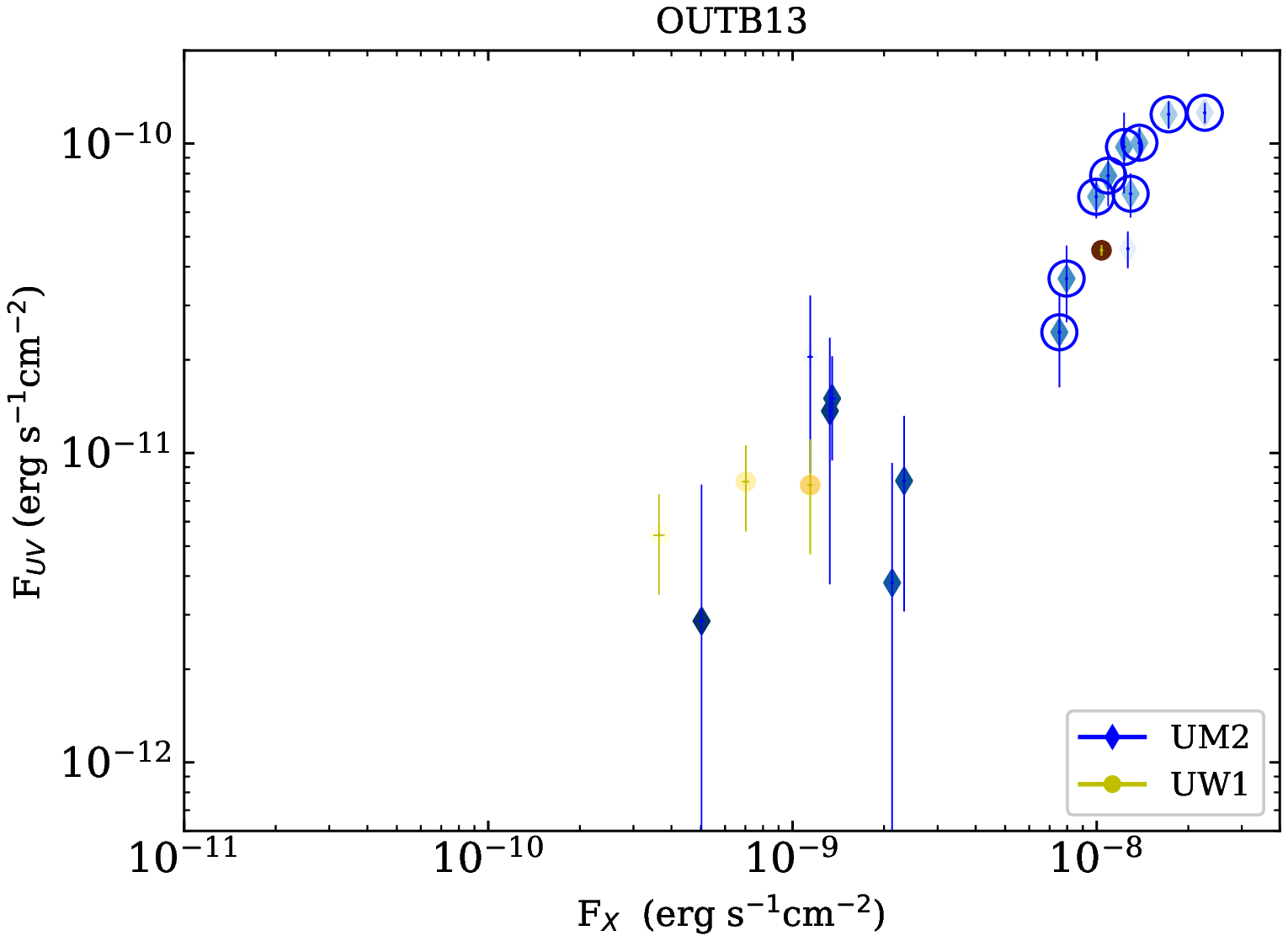}}
\subfigure{\includegraphics[width=0.8\columnwidth]{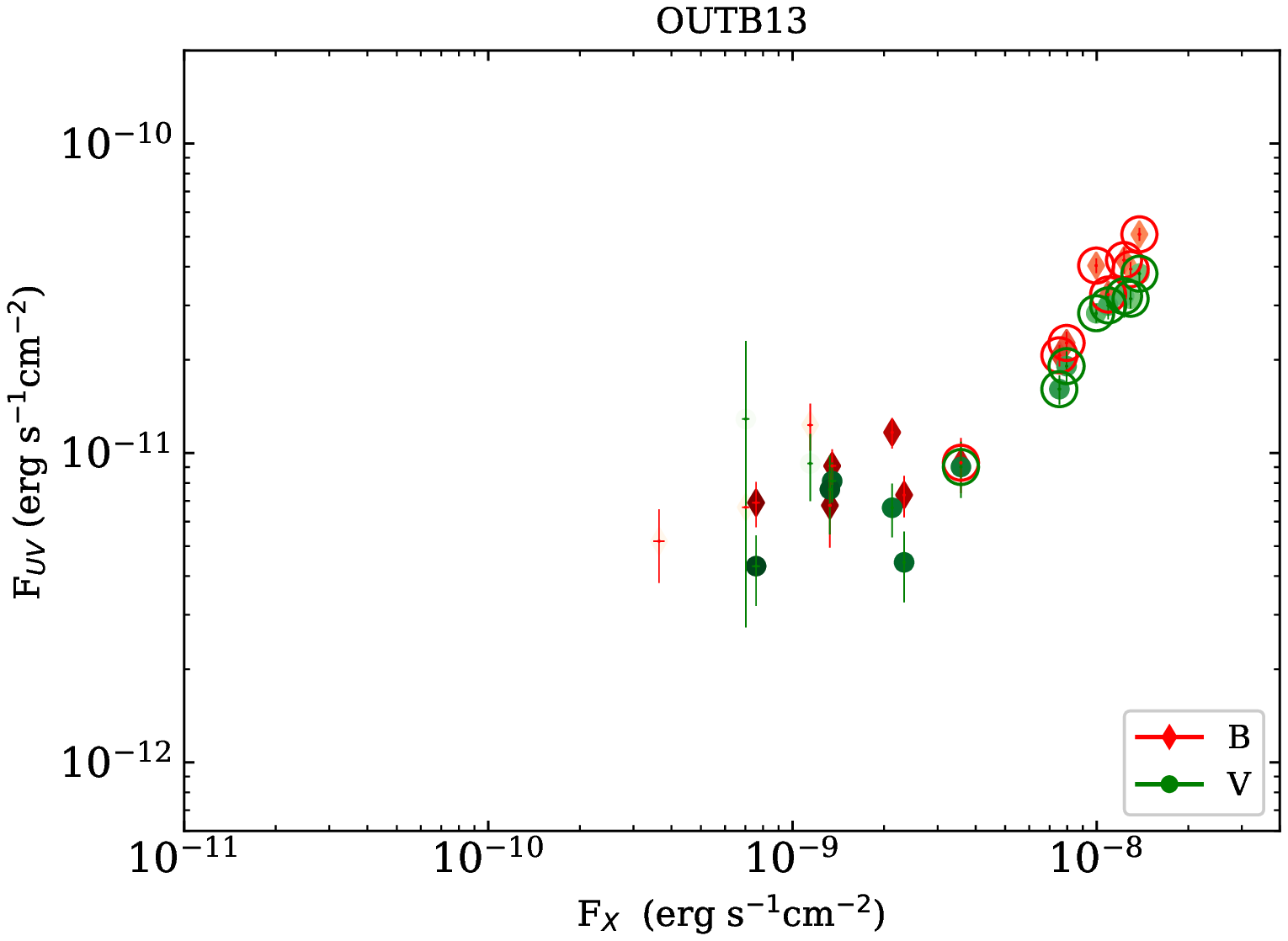}}
\subfigure{\includegraphics[width=0.8\columnwidth]{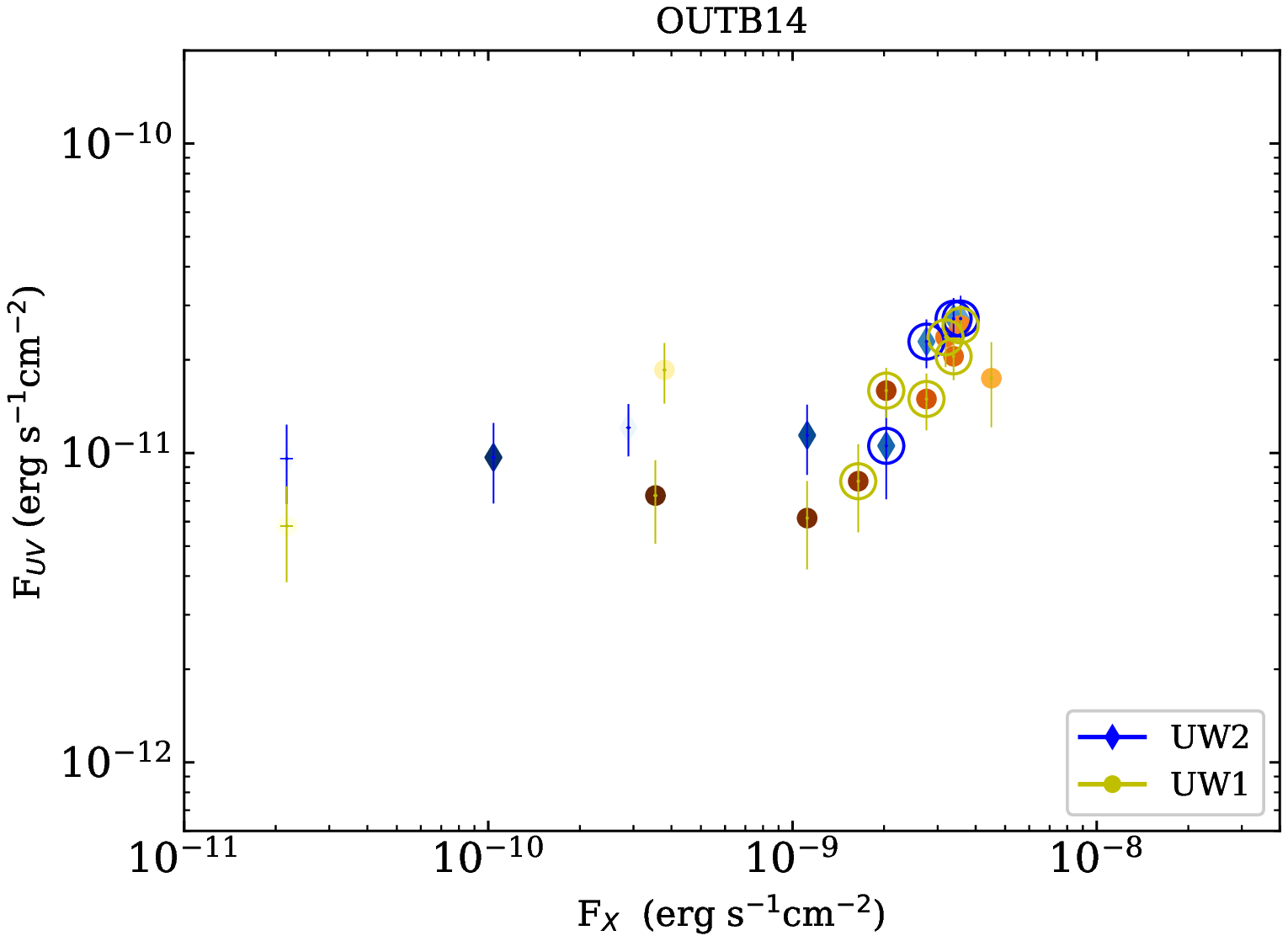}}
\subfigure{\includegraphics[width=0.8\columnwidth]{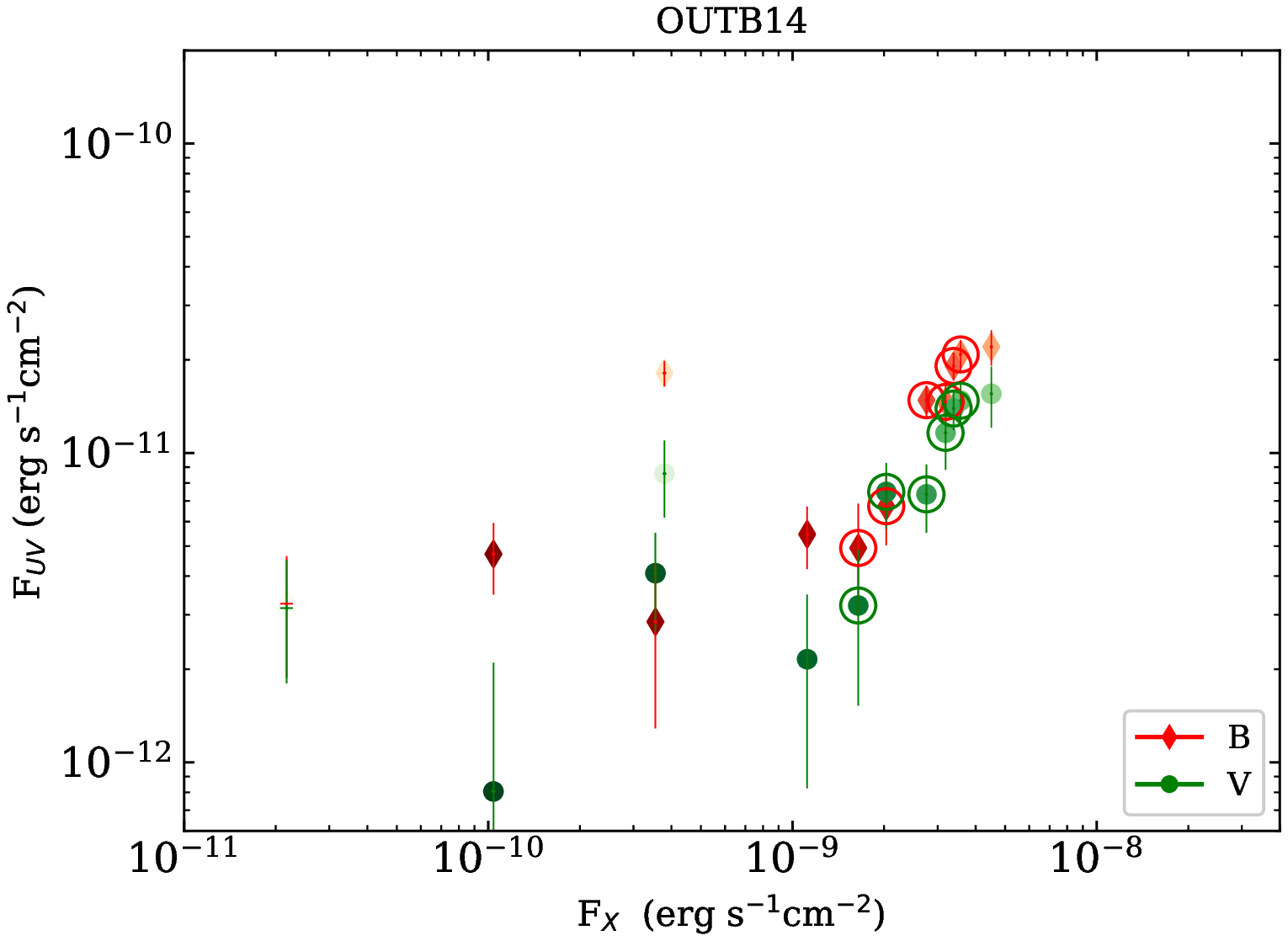}}
\subfigure{\includegraphics[width=0.8\columnwidth]{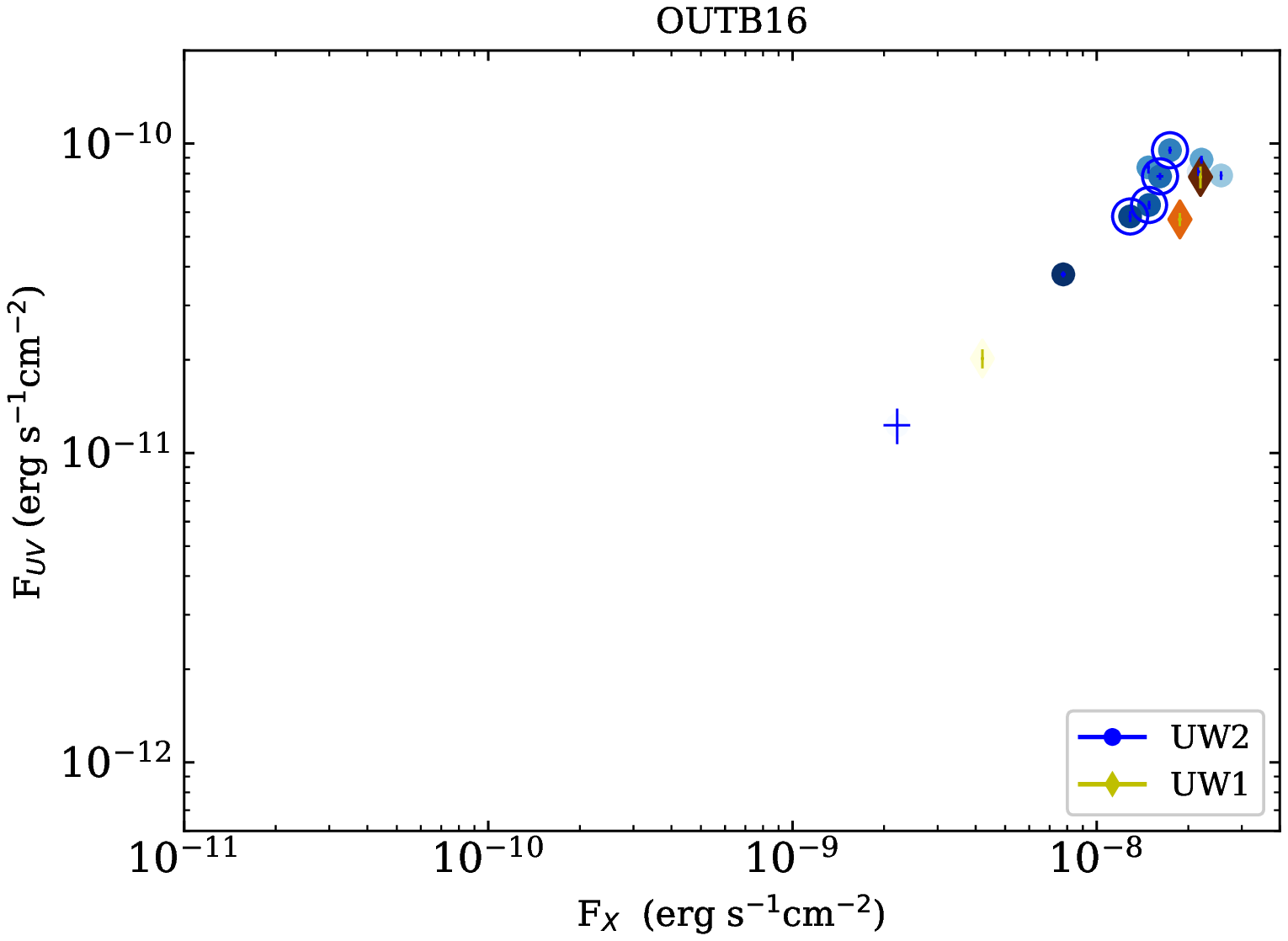}}

\caption{Evolution of the optical flux in the UVOT filters as a function of the 2--10 keV X-ray flux for the 2013, 2014 and 2016 outbursts. The colour of the markers reflects how far along the outburst the observations were taken, with a darker colour implying a later time. The markers used to fit the decay of the outbursts are circled in the plots. The plots are shown with the same axes limits to allow for a direct comparison.}
\label{fig:corr_filters}
\end{center}
\end{figure*}

In Figure \ref{fig:corr_filters} we show the plots of the UV/optical and X-ray flux, similar to Figure~\ref{fig:correlations}, but now for additional UVOT filters.

\section{Additional UV/optical and X-ray correlations fits}
\label{sec: appendixB}
\begin{table}
\caption{Results of the UV/optical and the 0.5-10 keV X-ray correlations. }

\resizebox{0.5\textwidth}{!}{%
\begin{threeparttable}
\begin{tabular}{ccccccc}

\hline\hline
&UVOT filters & \multicolumn{2}{c}{Rise} &\multicolumn{2}{c}{Decay}\\
  & (\#obs rise$/$decay)\tnote{a} & $\beta \pm \Delta \beta $\tnote{ab} & $\rho$ (p-value)& $\beta \pm \Delta \beta $\tnote{ab} & $\rho$ (p-value)\\

\hline 
\multirow{5}{*}{OUTB13}&UM2 (2/9)&--& --&1.14$\pm$0.20&0.95 (10$^{-5}$)\\
    &UW1 (4/--)&0.67$\pm$0.05&0.80(0.2)&--&--\\
    &\textbf{U (4/11)}&\textbf{0.60$\pm$0.03}&\textbf{ 1.0 (0.0)}&\textbf{1.49$\pm$0.10}&\textbf{0.97 (10$^{-7}$)}\\
    &B (3/8)&0.8$\pm$0.3& 1.0 (0.0)&1.32$\pm$0.20&0.88 (10$^{-3}$)\\
    &V (2/8)&--& --&1.20$\pm$0.13&0.98 (10$^{-5}$)\\
    \hline 
\multirow{5}{*}{OUTB14}&UW2 (2/4)&--& --&1.12$\pm$0.16&1.0 (0.0)\\
   &UW1 (3/6)&0.20$\pm$0.11&0.5 (0.67)&0.8$\pm$0.3&0.89 (10$^{-2}$)\\
    &\textbf{U (3/6)}&\textbf{0.32$\pm$0.04}&\textbf{1.0 (0.0)}&\textbf{1.17$\pm$0.19}&\textbf{0.83 (10$^{-2}$)}\\
    &\textbf{B (3/6)}&\textbf{0.27$\pm$0.15}&\textbf{1.0 (0.0)}&\textbf{1.33$\pm$0.13}&\textbf{0.94(10$^{-3}$)}\\
    &\textbf{V (3/6)}&\textbf{0.286$\pm$0.019}&\textbf{1.0 (0.0)}&\textbf{1.2$\pm$0.3}&\textbf{0.94 (10$^{-3}$)}\\
    \hline
\multirow{3}{*}{OUTB16}&UW2 (3/4)&0.76$\pm$0.16&0.5 (0.67)&1.8$\pm$0.4&1.0 (0.0)\\
    &UW1 (3/--)&0.76$\pm$0.10&1.0 (0.0)&--&--\\
    &U (1/7)&--& --&1.03$\pm$0.11&1.0 (0.0)\\
\hline
\end{tabular}
\begin{tablenotes}
  \item[a] The number of observations obtained during the rise/decay are given in parenthesis.
  \item[b] Errors reflect 1$\sigma$ confidence intervals.
  \item[c] We only attempted to fit correlations when 3 or more observations were available.
  \item[] The values indicated in bold highlight the filters/outbursts where the rise gives a significantly different correlation than the decay.
  \end{tablenotes}
\label{table:correlations05}
\end{threeparttable}
}

\end{table}
In Table \ref{table:correlations05} we list the results for fitting correlations to the UV/optical and 0.5--10 keV X-ray flux. The results are qualitatively similar to the results reported in the main body of the paper when using the 2--10 keV X-ray flux.

\bsp	
\label{lastpage}
\end{document}